\documentclass[journal]{IEEEtran}
\usepackage{bm}
\usepackage{graphicx} 
\usepackage{multirow}
\usepackage{algorithm}
\usepackage[noend]{algpseudocode}
\usepackage{amsmath}
\usepackage{graphics}
\usepackage{epsfig}
\usepackage{booktabs}
\usepackage{cite}
\usepackage{hyperref}
\usepackage{threeparttable}
\usepackage{xcolor}
\usepackage{colortbl}
\definecolor{mygray}{gray}{0.9}
\definecolor{mypink}{rgb}{.99,.91,.95}
\definecolor{mycyan}{cmyk}{.3,0,0,0}
\usepackage{caption}

\begin{document}
\title{Learned Fast HEVC Intra Coding}
\author{
        Zhibo~Chen,~\IEEEmembership{Senior~Member,~IEEE,}
        Jun~Shi
        and~Weiping~Li,~\IEEEmembership{Fellow,~IEEE}
\thanks{Zhibo Chen, Jun Shi and Weiping Li are with the CAS Key Laboratory of Technology in Geo-spatial Information Processing and Application System, University of Science and Technology of China, Hefei 230027, China. This work was supported in part by NSFC under Grant U1908209, 61632001 and the National Key Research and Development Program of China 2018AAA0101400. (e-mail: chenzhibo@ustc.edu.cn; shi1995@mail.ustc.edu.cn; wpli@ustc.edu.cn).}
}
\markboth{}
{Shell \MakeLowercase{\textit{et al.}}: Bare Demo of IEEEtran.cls for IEEE Journals}

\maketitle
\begin{abstract}
In High Efficiency Video Coding (HEVC), excellent rate-distortion (RD) performance is achieved in part by having a flexible quadtree coding unit (CU) partition and a large number of intra-prediction modes. Such an excellent RD performance is achieved at the expense of much higher computational complexity. In this paper, we propose a learned fast HEVC intra coding (LFHI) framework taking into account the comprehensive factors of fast intra coding to reach an improved configurable tradeoff between coding performance and computational complexity. First, we design a low-complex shallow asymmetric-kernel CNN (AK-CNN) to efficiently extract the local directional texture features of each block for both fast CU partition and fast intra-mode decision. Second, we introduce the concept of the minimum number of RDO candidates (MNRC) into fast mode decision, which utilizes AK-CNN to predict the minimum number of best candidates for RDO calculation to further reduce the computation of intra-mode selection. Third, an evolution optimized threshold decision (EOTD) scheme is designed to achieve configurable complexity-efficiency tradeoffs. Finally, we propose an interpolation-based prediction scheme that allows for our framework to be generalized to all quantization parameters (QPs) without the need for training the network on each QP. The experimental results demonstrate that the LFHI framework has a high degree of parallelism and achieves a much better complexity-efficiency tradeoff, achieving up to 75.2\% intra-mode encoding complexity reduction with negligible rate-distortion performance degradation, superior to the existing fast intra-coding schemes. 
\end{abstract}

\IEEEpeerreviewmaketitle

\section{Introduction}
\IEEEPARstart{T}{he} latest video coding standard, High Efficiency Video Coding (HEVC) \cite{6316136}, developed by the Joint Collaborative Team on Video Coding (JCT-VC) in 2012, improves the coding performance significantly. Compared with the previous video coding standard, H.264/AVC \cite{1218189}, HEVC saves about 50\% bits with the same perceptual video quality using several elaborate methods. Specifically, for intra coding \cite{6317153}, the coding unit (CU) is recursively divided into a quadtree-based structure from the largest CU 64$\times$64 to the smallest CU 8$\times$8. In addition, up to 35 intra-prediction modes for the prediction unit (PU) are allowed. These two techniques are beneficial to the coding performance, with the expense of an enormous complexity increase, which makes it intractable for some real-time applications. Therefore, there is a need for complexity reduction of HEVC intra coding.

The past few years have witnessed many fast algorithms for HEVC intra coding, which can be classified into two main categories: fast CU size decision and fast intra-mode decision. For CU size decision, due to the flexibility of CU size in HEVC, the recursive process of the quad-tree CU partition will bring tremendous complexity. Many approaches try to predict the CU partition pattern in advance; thus, the brute-force recursive RD optimization (RDO) search can be avoided. Some heuristic methods \cite{6983560,6470665,zhao2014fast,7538911,7521923,zhang2019statistical,6738325,6890319, chiang2019fast} exploit the intermediate characteristics of current CU and the spatial correlations with the neighboring CUs to early determine the coding tree unit (CTU) partition result. Specifically, the approaches in \cite{zhang2019statistical,6738325,6890319, chiang2019fast} extract texture features of CU to determine the CU size. The Hadamard (HAD) cost and RD cost are utilized in \cite{6983560,6470665,zhao2014fast} to execute early CU split and early CU termination. Kim \textit{et al.} \cite{7521923} proposed splitting a CU based on the number of high-frequency key points. Recently, several machine learning approaches \cite{7457241,7024895,kuanar2019adaptive,7588907,duanmu2016fast,westland2019decision,kuang2019online,7547305,8019316,chung2017hevc,8384310} have been proposed for fast CTU partition. In \cite{duanmu2016fast, westland2019decision}, decision trees are trained for early termination decisions. Zhang \textit{et al.} \cite{7457241} proposed a CU depth decision method based on support vector machine (SVM). Bayesian decision rules are utilized in \cite{7024895,kuang2019online} to decide the CU size. Since 2016, CNN has been studied for CTU partition due to its ability to automatically extract features for CTU structure determination. Liu \textit{et al.} \cite{7547305} proposed a VLSI friendly algorithm with a shallow CNN structure for CTU partition, and Xu \textit{et al.} \cite{8384310} developed a deep CNN-based approach to predict the CTU partition. Kuanar \textit{et al.} \cite{kuanar2019adaptive} proposed using a CNN to detect textures and object shapes in CUs and then classify the spatial patterns into four classes for CU depth prediction.

 As for the intra-mode decision, it leads to much higher complexity if each intra-mode performs an RDO calculation. The current HEVC encoder has already adopted a three-step algorithm to expedite the process of intra-mode decision \cite{piao2010encoder}. In the first step, rough mode decision (RMD) is used to select several candidate modes with the least Hadamard transform-based costs (HAD costs). Second, the most probable modes (MPMs) are added to the candidate list. Finally, all the modes in the candidate list go through RDO, which requires higher complexity, to find the best mode. In the past few years, many approaches \cite{6201851,6466835,7169266,7149261,7024895,6662471,7805540,shen2013fast,7362704,7457241,7051618,8412615, kuanar2019adaptive, chiang2019fast} have been proposed for fast intra-mode decisions to further reduce the complexity. Specifically, texture and edge information was extracted in \cite{6466835} and \cite{7149261} to predict the possible best intra-prediction mode. Hu \textit{et al.} \cite{7024895} applied a fast intra-mode decision algorithm based on the outliers with entropy coding refinement. For the mode searching procedure, a progressive rough mode searching technique was proposed in  \cite{6662471}. Similarly, Liao \textit{et al.} \cite{7805540} adjusted the order of MPM and RMD and then used the depth of current PU to choose the most probable modes. Jaballah \textit{et al.} \cite{7362704} proposed clustering the set of intra modes into groups and selectively choosing the candidates for RDO calculations. Laude \textit{et al.} \cite{laude2016deep} used CNN models to directly predict the optimal mode to avoid extra RDO calculations. \cite{7457241} proposed a gradient-based fast algorithm, calculating the average gradients in the vertical direction and horizontal direction for every PU, to reduce the number of candidate modes. Ryu \textit{et al.} \cite{8412615} adopted the random forest algorithm to estimate the possible intra-prediction modes. Most of these methods tend to discover the relationship between the mode and the content attributes and then heuristically detect special content statistical features (such as homogeneous blocks or vertical texture blocks) to simplify the RMD or RDO procedures. 
 
In general, the aforementioned methods have made good explorations on fast intra coding from different specific perspectives and have achieved good performance improvements. However, there are still many comprehensive factors of fast intra coding to be considered, such as the efficiency of feature extraction, the validity of the best candidate list, the tradeoff between complexity and coding performance, the impact of different quantizations, and parallelism for easy hardware implementation. 

Therefore, with comprehensive consideration of the above factors, we propose a learned fast HEVC intra coding (LHFI) framework in this paper. A special low-complex asymmetric-kernel CNN (AK-CNN) is designed to efficiently extract near-horizontal and near-vertical textures, which are important patterns for intra-coding mode prediction. For fast CU/PU size decision, our AK-CNN can perform the decisions of early split and early termination. For fast intra-mode decision, we introduce the concept of minimum number of RDO candidates (MNRC) and use the AK-CNN to predict the number of valid best RDO candidates for every PU adaptively. To provide the configurable trade-offs between the complexity reduction and coding performance, we adopt the confidence threshold scheme and then use the evolutionary algorithm to explore the optimal combination of the threshold values. We also propose a novel interpolation-based prediction scheme for LFHI framework to achieve its generalization capability to variant quantization cases. Finally, it is important to note that neighboring reconstructed pixels are not required for the LFHI framework, which guarantees LFHI's parallelism for friendly hardware implementation. Our experimental results demonstrate that the LFHI framework has a high degree of parallelism and better complexity-efficiency tradeoff and can reduce the encoding time of HEVC by 75.2\%, with a negligible 2.09\% Bj$\phi$ntegaard delta bit-rate (BD-BR) increase, over the JCT-VC test sequences. The performance significantly outperforms other state-of-the-art approaches. In brief, the main contributions of this paper are summarized as follows:

\begin{itemize}
	\item We design a novel AK-CNN structure for effective texture feature extraction, which can be used for both fast CU/PU size decision and intra-mode decision.
	\item We propose a solution for fast intra-mode decision from a new perspective. A significant attribute of every PU: MNRC is introduced. Based on the prediction of MNRC, we can remove the redundant RDO candidates in a safer manner.
	\item We propose an evolution optimized threshold decision (EOTD) scheme to explore the optimal configurable complexity-efficiency tradeoffs for HEVC intra coding.
	\item With the proposed interpolation-based prediction scheme, our framework is generalized to all quantization parameters (QPs) superbly.
\end{itemize}

The remainder of this paper is organized as follows. Section II introduces the overview of HEVC intra-coding scheme. In Section III, we introduce the proposed LFHI framework. The detailed schemes for fast CU/PU size decisions and fast intra-mode decisions are described in Section IV and Section V, respectively. We present the experimental results and analysis in Section VI and then conclude the paper in Section VII. We will release the code and dataset later online at \url{http://staff.ustc.edu.cn/~chenzhibo/resources.html} for public research usage.

\section{Overview of HEVC Intra Coding}
\begin{figure}[h]
	\centerline{\includegraphics[scale=0.28]{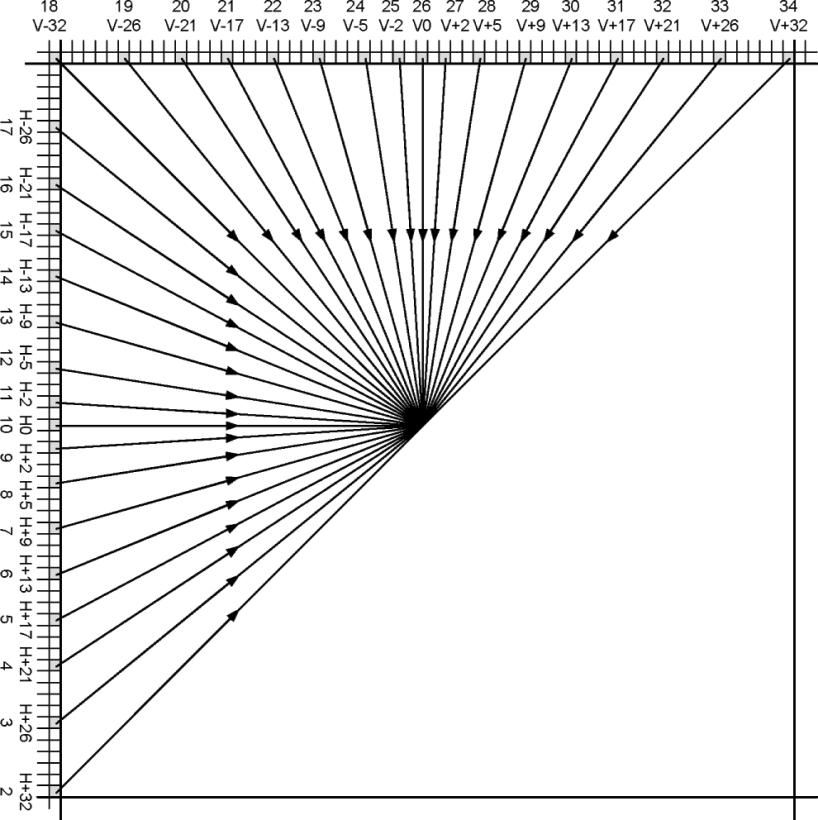}}	
	\caption{Angular intra-prediction modes. H and V indicate the horizontal and vertical directionalities, respectively\cite{6317153}.}
	\label{intra}
\end{figure}

In HEVC intra coding, the CU sizes can be 64$\times$64, 32$\times$32, 16$\times$16 and 8$\times$8, which correspond to depths 0$\sim$3. For the smallest CU 8$\times$8, two different PU sizes (8$\times$8 and 4$\times$4) are possible. The PU size 4$\times$4 indicates that there are four PUs of equal size in that CU. In this paper, the PU 4$\times$4 is viewed as depth 4. For every PU, up to 35 intra-prediction modes, including planar, DC and 33 angular modes, are allowed. Fig. \ref{intra} shows the prediction directions associated with the angular modes. It requires extremely high computational complexity to execute the RDO calculation for all modes.

\begin{figure}[]
	\centerline{\includegraphics[scale=0.9]{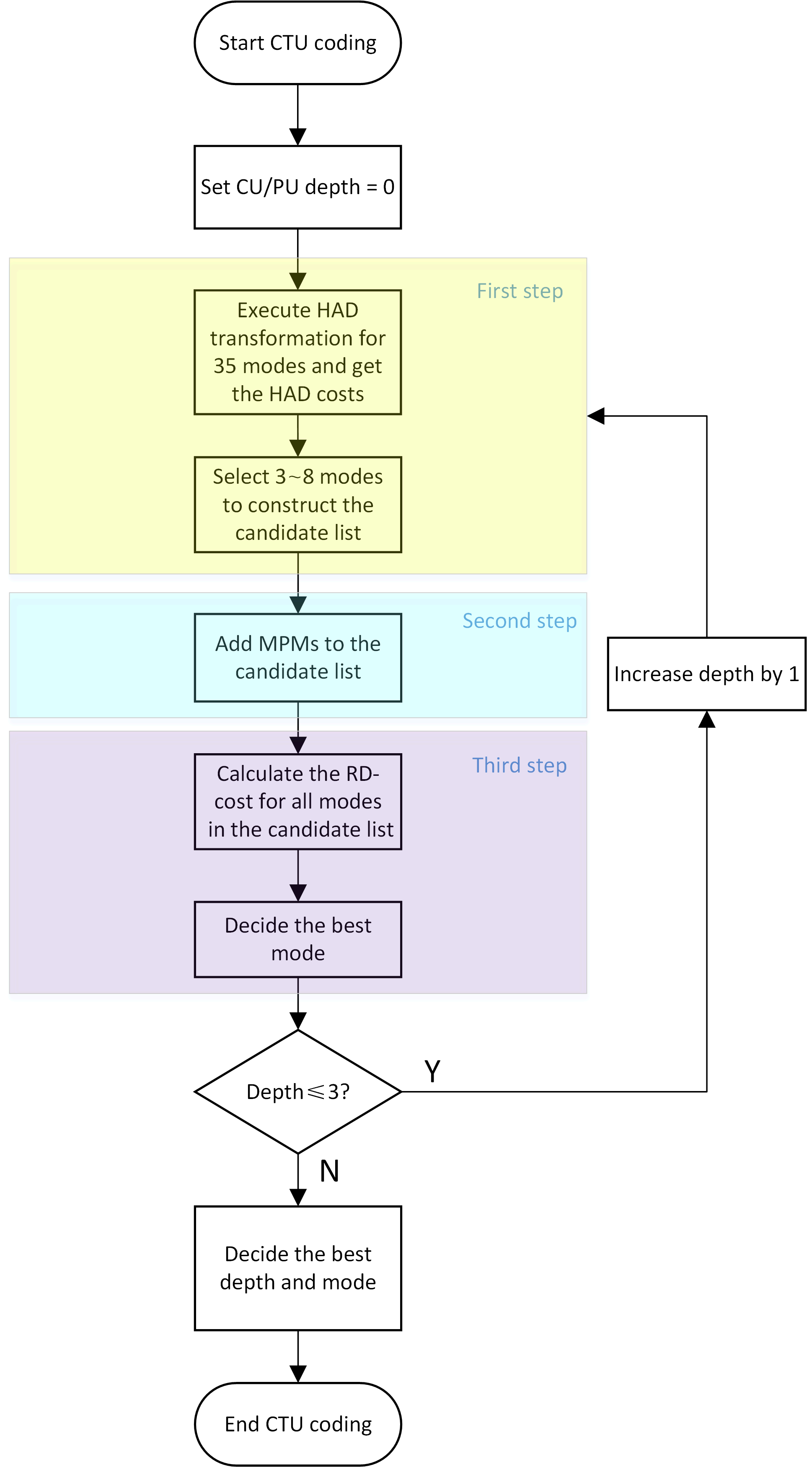}}
	\caption{HEVC intra-coding flowchart.}
	\label{flow1}
\end{figure}

The flowchart of the encoder of HEVC reference software HM 16.9 \cite{HM} is shown in Fig. \ref{flow1}. As we can see, the encoder already adopts a three-step intra-mode decision fast algorithm. In the first step, RMD, a low-complexity procedure, is designed to construct a candidate list for RDO based on the Hadamard transform-based cost ($J_{HAD}$), which is calculated by
\begin{equation}
J_{HAD}=SATD+\lambda\cdot R_{mode}
\end{equation}
where $SATD$ represents the sum of absolute transformed difference. $\lambda$ is the Lagrangian multiplier decided by the QP, and $R_{mode}$ is the number of bits to encode the information of intra-prediction mode. After the above calculations, three best modes with the least $J_{HAD}$ are selected for PUs with the sizes 64$\times$64, 32$\times$32 and 16$\times$16, while eight optimal modes are chosen for PUs with the sizes 8$\times$8 and 4$\times$4. In the second step, the three modes generated from the modes of neighboring PUs, \textit{i.e.}, the MPMs, are added to the candidate list. In the last step, all the modes in the list go through RDO to find the best intra-prediction mode based on the RD-cost ($J_{RDO}$), which is computed by
\begin{equation}
J_{RDO}=SSE+\lambda\cdot R_{total}
\end{equation}
where $SSE$ denotes the sum of the squared errors and $R_{total}$ is the number of total bits used to encode the CU. 

If the depth of current CU does not reach the largest depth, the depth is increased by 1, and four sub-CUs continue to perform the above process. For the smallest CU, two different PU sizes (8$\times$8 and 4$\times$4) are tested.
 
This three-step algorithm reduces a large number of intra coding calculations in the HEVC encoder. However, as we mentioned in the introduction section, there is still large space for further improvement.

\section{Proposed Framework}
\begin{figure}[h]
	\centering
	\centerline{\includegraphics[scale=0.9]{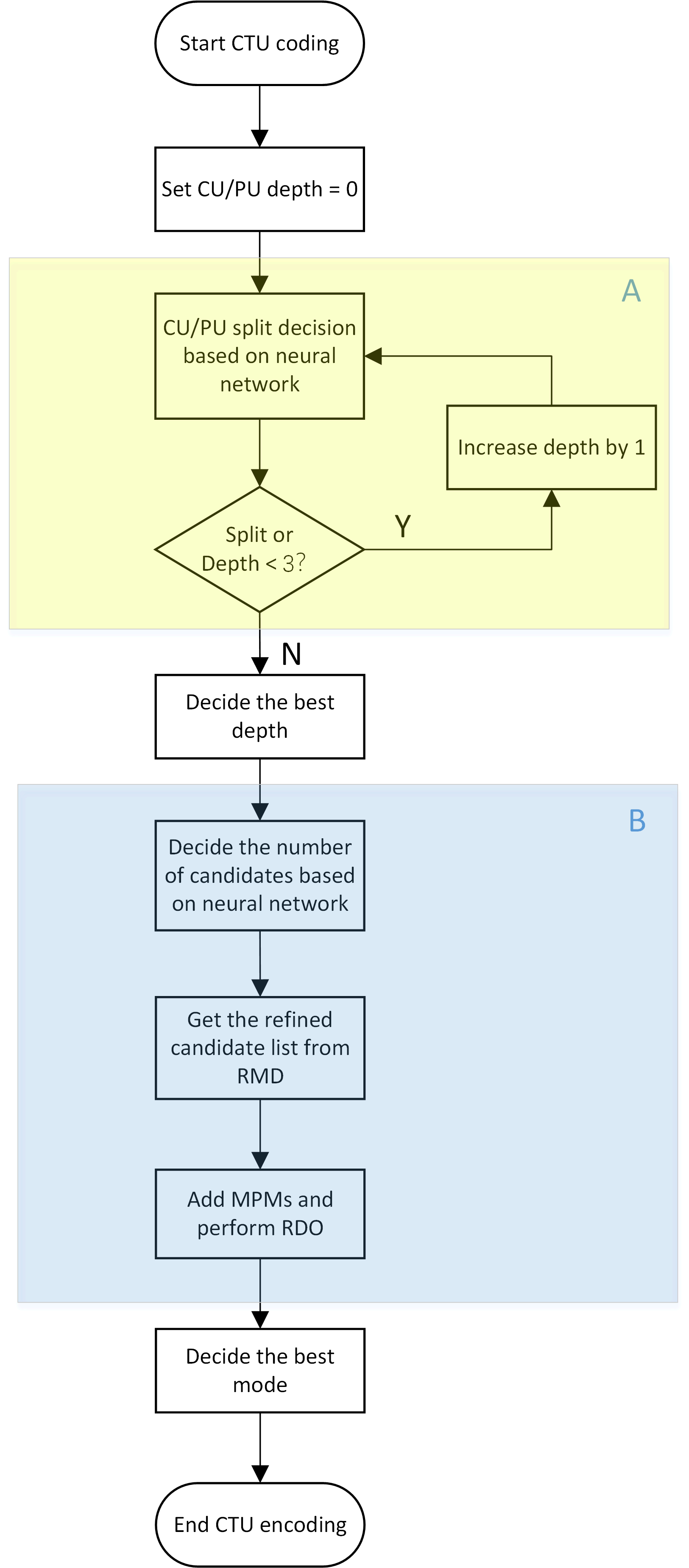}}
	\setlength{\belowcaptionskip}{0pt}
	\caption{Flowchart of the proposed LFHI framework. A: fast CU/PU size decision. B: fast mode decision.}
	\label{flow2}
	\vspace{-0.1cm}
\end{figure}

Fig. \ref{flow2} shows the flowchart of our proposed LFHI framework for fast intra coding, which aims at reducing the computational complexity in CU/PU size decision and intra-mode decision while maintaining the RD performance. Unlike other approaches that make use of the reconstructed pixels or some intermediate variables such as RD-cost, our framework only uses the pixels in the current CU; thus, it can be implemented as preprocessing before encoding. As a consequence, our scheme has a high degree of parallelism.

 For fast CU/PU size decision (Block A in Fig. \ref{flow2}), we employ AK-CNNs as classifiers in four depths. The classifiers output the splitting decision. The CU/PU is split unless the decision is nonsplitting or it reaches the maximal depth. 

As for the fast intra-mode decision (Block B in Fig. \ref{flow2}), for PU in every depth, the AK-CNNs decide the number of best candidate modes for RDO calculation adaptively, which enables the RMD to select a flexible number of modes.

After above preprocessing operations, the encoder ensures the final CU/PU size and intra-mode RDO modes according to this information. The details for fast CU/PU size decision and fast intra-mode decision are described in the following sections.

\section{Fast CU/PU-Size Decision}
\subsection{Four-Level Classifier}
In HEVC intra coding, the best CU/PU size is decided by searching all the CU/PU partition pattern with RDO calculation, which means that every CU/PU is encoded for several times, resulting in redundancy and enormous computational complexity. If we can construct a classifier that can determine the CU/PU size in advance, large encoding computational complexity can be reduced. 

\begin{figure*}[h]
	\centerline{\includegraphics[scale=0.98]{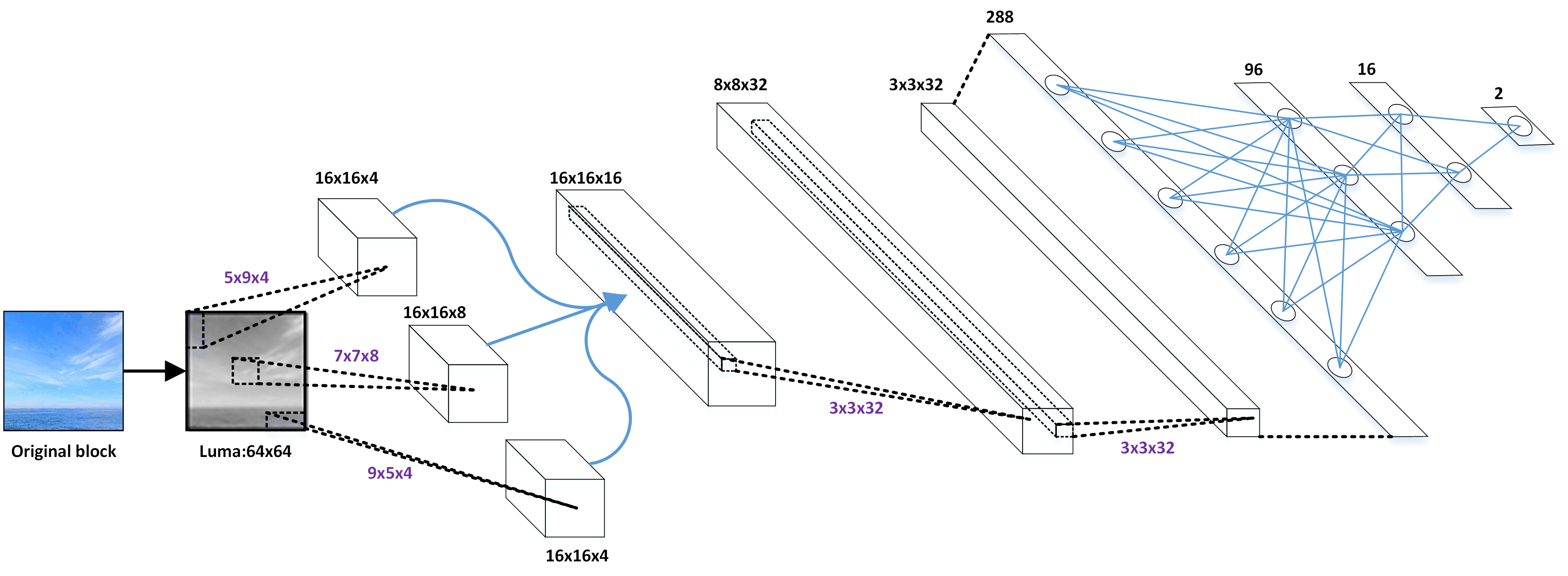}}	
	\caption{Structure of AK-CNN. Taking the luminance of the current CU as input, the AK-CNN will output the splitting decision. Closer attention will be paid to the near-horizontal and near-vertical textures.}
	\label{ak}
	\vspace{-0.1cm}
\end{figure*}

Note that the entire CTU partition classification problem can be viewed as a combination of four-level binary classifications. Due to the large number of modes of the CTU partition, direct prediction of the partition mode is inaccessible and resource-wasting. Therefore, we try to design a scheme with separate classifiers at four decision levels.

\subsection{AK-CNN Structure}
Some previous approaches attempted traditional machine learning tools (such as SVM) to implement such classifiers. Several manual features, such as texture strength and local variance, are extracted. However, since the results are heavily dependent on the design of the features, these methods introduce limitations and lack generalization in certain specific situations. Recently, CNNs have achieved dominant performance in many vision tasks, such as classification and segmentation. CNNs' outstanding feature extraction and learning capabilities make them extremely competitive.

In this paper, we propose a special AK-CNN as the classifier to take full advantage of the breadth of neural networks. Taking the luminance of current CU as input, the AK-CNN outputs a splitting decision. Since we need four classifiers to decide the CU/PU size, we build four CNNs to implement the above classifiers at each level of the selected QP. Due to the low complexity requirement, a relatively shallow and light architecture is adopted in this paper.

For this particular task, we design a novel asymmetric-kernel structure. As shown in Fig. \ref{intra}, to reflect the statistical prevalence of angles and the effectiveness of signal prediction processing, the modes are intentionally designed to provide denser coverage of near-vertical and near-horizontal angles \cite{6316136}. As a result, the textures of near-vertical and near-horizontal are more important for the intra prediction, which is also significant for the partition result. Therefore, it is necessary and meaningful for the CNN to pay more attention to such texture patterns. To this end, our neural network extends the first convolutional layer to three different branches. The first branch is the traditional convolutional layer, with normal square kernels, while the remaining two branches have convolutional kernels of asymmetric shape that target at detecting near-vertical or near-horizontal textures. This structure could enable the neural network to detect the texture features more effectively and efficiently. All three branches output equal-sized feature maps. Then, we concatenate them in depth and put them into two convolutional layers with small kernels to learn the correlation between these features. The combination of these features could help the CNN better understand the content characteristics. Next, the extracted features flow through three fully connected layers to get the final prediction. In our neural network, the activation function of all convolutional layers and hidden fully connected layers is  leaky rectified linear unit (LeakyReLU) with $\alpha = 0.25$, while the output layer is activated with Softmax function. Fig. \ref{ak} presents the details of AK-CNN for block 64$\times$64, and the information about other AK-CNNs is given in Table \ref{tab-ak}.

\begin{table}[!h]
	\renewcommand{\arraystretch}{1.25}
	\centering
	\caption{Proposed AK-CNN structure}	
	\begin{tabular}{|c|c|c|c|c|c|}
		\hline
		\multicolumn{6}{|c|}{AK-CNN Configuration}                                                                                                                                                                                                                                                                                                                                                                      \\ \hline
		\multicolumn{3}{|c|}{A}                                                                                                                                                                           & \multicolumn{3}{c|}{B}                                                                                                                                                                                 \\ \hline
		\begin{tabular}[c]{@{}c@{}}conv\\ 7x3 -4\\ stride:2\end{tabular} & \begin{tabular}[c]{@{}c@{}}conv\\ 5x5 -8\\ stride:2\end{tabular} & \begin{tabular}[c]{@{}c@{}}conv\\ 3x7 -4\\ stride:2\end{tabular} & \begin{tabular}[c]{@{}c@{}}conv\\ 5x1 -4\\ stride:1\end{tabular} & \begin{tabular}[c]{@{}c@{}}conv\\ 3x3 -8\\ stride:1\end{tabular} & \begin{tabular}[c]{@{}c@{}}conv\\ 1x5 -4\\ stride:1\end{tabular} \\ \hline
		\multicolumn{3}{|c|}{concatenate}                                                                                                                                                                      & \multicolumn{3}{c|}{concatenate}                                                                                                                                                                       \\ \hline
		\multicolumn{3}{|c|}{\begin{tabular}[c]{@{}c@{}}conv 3x3 -32\\ stride:2\end{tabular}}                                                                                                                & \multicolumn{3}{c|}{\begin{tabular}[c]{@{}c@{}}conv 3x3 -32\\ stride:2 (1) \end{tabular}}                                                                                                              \\ \hline
		\multicolumn{3}{|c|}{\begin{tabular}[c]{@{}c@{}}conv 3x3 -32\\ stride:2\end{tabular}}                                                                                                                & \multicolumn{3}{c|}{\begin{tabular}[c]{@{}c@{}}conv 3x3 -32\\ stride:2\end{tabular}}                                                                                                                 \\ \hline
\multicolumn{3}{|c|}{$FC_1$: 96}                                                                                                                                                                            & \multicolumn{3}{c|}{$FC_1$: 96}                                                                                                                                                                             \\ \hline
\multicolumn{3}{|c|}{$FC_2$: 16}                                                                                                                                                                            & \multicolumn{3}{c|}{$FC_2$: 16}                                                                                                                                                                             \\ \hline
\multicolumn{3}{|c|}{$FC_3$: 2}                                                                                                                                                                             & \multicolumn{3}{c|}{$FC_3$: 2}                                                                                                                                                                              \\ \hline
	\end{tabular}
\begin{tablenotes}
	\item Network A is for block 32$\times$32, while network B is for 16$\times$16 and 8$\times$8.
\end{tablenotes}
\vspace{-0.cm}
\label{tab-ak}
\end{table}
Based on the AK-CNN, redundant RDO calculations are avoided, which could lead to a large computational complexity reduction. 

\subsection{Evolution Optimized Threshold Decision}
\begin{figure}[!h]
	\centerline{\includegraphics[scale=.168]{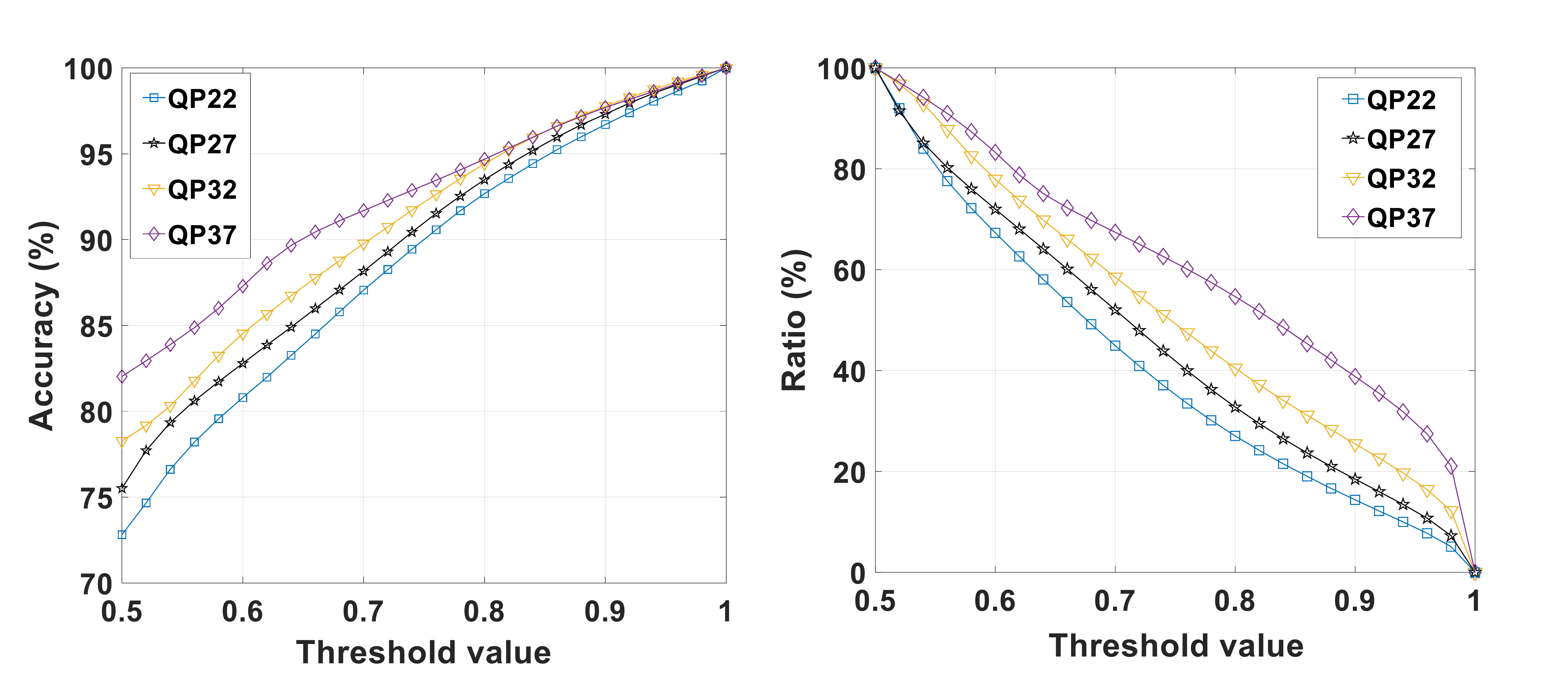}}
	\caption{Relationship between threshold and accuracy/ratio in depth 3 in our validation set.}
	\label{th}
\end{figure}

For any classifier, it could not be absolutely accurate under all circumstances, and a false prediction will result in coding performance degradation. Therefore, we adopt a confidence threshold scheme to ameliorate this issue. For AK-CNN, the output Softmax value of the chosen action (splitting or nonsplitting) can be considered as the confidence for this classification trial. In general, the larger the value, the more confident the classifier is. Therefore, via setting the threshold of Softmax value, we can decrease the false prediction rate in the subsets with large Softmax values, which we define as the confident classifications. Classification results with uncertainty (do not pass the threshold) will not be adopted. For those blocks, we will check them with RDO calculations as normal. Fig. \ref{th} shows the results of prediction accuracy and ratio of confident classifications with the change of threshold in four QPs in depth 3 (from PU 8$\times$8 to PU 4$\times$4) in our validation set.

The value setting of threshold is important, since a large threshold leads to high prediction accuracy (small coding performance degradation), while the computational complexity reduction rate is small, considering that many blocks tend to be checked. By setting the different thresholds, we can achieve a scalable complexity reduction scheme. However, how to achieve the best RD-complexity curve remains questionable. As we need to set four thresholds for each depth, it is actually a multiple-variable optimization problem. The complexity reduction rate depends on the ratio of fast decision, and the RD performance is dependent on the prediction accuracy, both of which rely on the thresholds. Therefore, we can assume the complexity reduction rate $C$ and RD performance degradation $R$ satisfy the following:
\begin{equation}
\begin{cases}
& C = f_1(th_1,th_2,th_3,th_4) \\
\\
& R = f_2(th_1,th_2,th_3,th_4)\
\end{cases}
\end{equation}
where $th_{i}$ denotes the threshold value at depth $i$. Our goal is to achieve higher complexity reduction rate $C$ and lower performance degradation $R$. To reach this target, we need to find a group of combinations of the threshold values. Each combination of thresholds $th$ in this group should satisfy the following expression for all other $th'$:
\begin{equation}
C(th)>C(th')\quad \quad \quad if\ R(th)\geq  R(th').
\end{equation}
Such $th$ point is called a \textit{Pareto optimal point}\cite{hochman1969pareto}. Any improvement in a Pareto optimal point in one objective has to result in deterioration in the other objective \cite{miettinen2012nonlinear}.

The aforementioned joint optimization of the complexity reduction rate and RD performance by adjusting four threshold values is a really challenging task. To solve such a problem, we introduce the evolutionary algorithm (EA), which is naturally good at dealing with noncontinuous and nonconvex optimization problems. Specifically, we adopt the multi-objective evolutionary algorithm based on decomposition (MOEA/D\cite{zhang2007moea}). It decomposes an optimization problem into several scalar subproblems, which could save the computation and be used to deal with disparately scaled objectives. 

In this paper, multiple chromosomes, representing distinct threshold values, are evolved simultaneously subject to complexity reduction rate $C$ and RD performance degradation $R$ as two adversarial objectives, which are the two evaluation metrics of this task. 

During the iterations of evolutionary algorithm, we set the range of threshold values to satisfy the following condition to obtain a solid result based on accuracy $acc$:
\begin{equation}
80\% \leq acc(th_i) \leq 98\% \quad \quad  for\ 0\leq i \leq 3.
\end{equation}
The initial population of our evolutionary algorithm is generated from uniform sampling, which denotes several combinations of threshold values. Then, we allocate them with weight vectors $\lambda_j$ to set up $N_{sub}$ subproblems. In the crossover phase, we adopt the differential evolution operator to generate new individuals. We also perform point mutation to modify the newly generated individuals. After that, the complexity reduction rate $C$ and RD performance degradation $R$ are calculated as the evaluation indicator of each individual. If the individual among $m_z$ points is not a dominated variable, then we update both the object and output. Tchebycheff aggregation function \cite{miettinen2012nonlinear} is used as the scalar optimization function $g^{te}$. Algorithm \ref{alg} shows the details of the whole iterations of EOTD.

\begin{algorithm}[h] 
	\caption{Evolution-based Approach for Threshold Values} 
	\begin{algorithmic}[1] 
		\Require 
		\begin{itemize}
			\item $f(x)$: multi-objective functions;
			\item $SC$: stopping criterion;
			\item $V$: validation set;
			\item $g^{te}$: scalar optimization function.
		\end{itemize}
		\Ensure Pareto set of threshold values.
		\State  \textbf{Step 1 Initialization:}  
		\State  Initialize the population $x$, compute $f(x)$ in $V$, and obtain the reference point $z$.
		\State  \textbf{Step 2 Update:}  
		\For{$i = 1,2,$...$,N_{sub}$} 
		\State  \textbf{Step 2.1 Reproduction and mutation: }  
		\State Using the differential evolution operator to generate a 
		\State new  combination of threshold values $x'$, and then, \State apply point mutation on it
		\State  \textbf{Step 2.2 Update of $z$:} 
		\For{$j = 1,2,$...$,m_{z}$} 
		\State if $f_j(x')<z_j$, then set $z_j=f_j(x')$.
		\EndFor 
		\State  \textbf{Step 2.3 Update of neighborhood solutions:} 
		\For{each neighborhood index $j$}
		\State if $g^{te}(x|{\lambda}_{j},z)\leq g^{te}(x_j|{\lambda}_{j},z)$, 
		\State then set $x_j = x$ and $f(x_j) = f(x)$.
		\EndFor
		\EndFor 
		\State  \textbf{Step 3 Stopping condition:} 
		\State If $SC$ is satisfied or the boundary is reached, then stop and output $x$. Otherwise, go to \textbf{Step 2}.
	\end{algorithmic} 
\label{alg}
\end{algorithm}

\subsection{Variant QP Adaption}
Due to the requirement of variant QP settings in the video encoding configuration, our fast algorithm also needs to adapt to the variant QPs. Previous CNN-based approaches (such as \cite{8384310}) always take the QP as an input of the neural network and use one model to adapt to all QPs. However, this leads to two main problems:

\begin{enumerate}
	\item \textbf{Large complexity}: using one CNN to learn the relationship between the partition patterns and the blocks for all QPs precisely needs a higher number of parameters in the network (high learning ability), which brings high complexity in every single classification trial.
	\item \textbf{Low precision}: it is difficult for such a single model to learn the partition patterns for every QP precisely. Instead, the model tends to predict the common results among different QPs. Therefore, the prediction precision declines, resulting in large coding performance degradation. Furthermore, most of the existing methods only use the training data of QPs in common settings, like \{22,27,32,37\}, as input. As a consequence, for the data not in this set (\textit{e.g.}, QP = 25), the CNN cannot predict it well.
\end{enumerate}

\begin{figure}[]
	\vspace{-0.2cm}
	\centerline{\includegraphics[scale=.14]{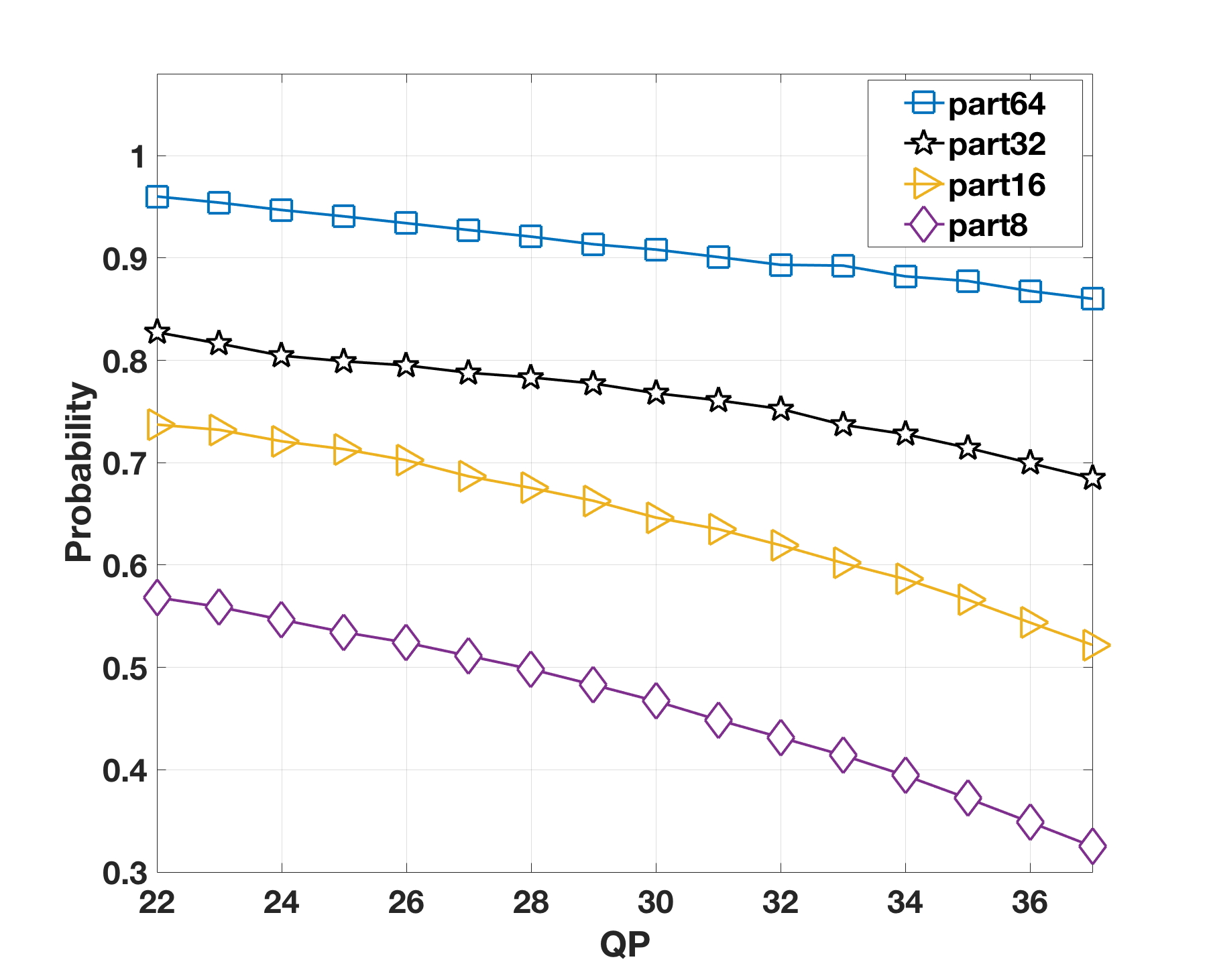}}
	\caption{Splitting rate among different QPs.}
	\label{split}
\end{figure}

In this paper, we introduce a novel scheme to employ four light CNNs to cover the variant QP range, instead of one large model. The basic idea is to use the models of the two nearest QPs to cooperatively predict the partition patterns for the chosen QP. First, we train four light AK-CNNs at the anchor QPs, \textit{i.e.}, QPs in \{22,27,32,37\}. For these QPs, the AK-CNNs output the partition result precisely with low computational complexity. Then, we use these four AK-CNNs to generalize to other QPs based on the relationship between the splitting rates among different QPs. We denote the splitting rate of QP $q$ in depth $i$ as $p_{i}^{q}$, which indicates the possibility of a block in depth $i$ to be further split. It is calculated as follows:
\begin{equation}
p_{i}^{q}= \frac{\sum_{k=i+1}^{4}S_k}{\sum_{k=i}^{4}S_k} \quad when\ QP=q
\end{equation}
where $S_k$ is the sum of pixel numbers of all blocks in depth $k$. The splitting rate in our validation set is shown in Fig. \ref{split}, from which we can observe that the splitting rate of every QP can be a linear combination of the splitting rates of the two nearest QPs.

Furthermore, the prediction probability vector (Softmax value) yielded from the AK-CNNs can also be considered as a special form of splitting rate for a specific block. Therefore, we can obtain the prediction probability vectors for other QPs by combining the vectors of the anchor QPs. The combination coefficients can be calculated from the equation set of the splitting rate:
\begin{equation}
\begin{cases}
& p_{i}^{q} = a_{i}^{q}\cdot p_{i}^{m}+b_{i}^{q}\cdot p_{i}^{n} \\
\\
& a_{i}^{q} + b_{i}^{q} = 1
\end{cases}
\end{equation}
where $m$ and $n$ are the two nearest QPs of $q$. By solving this equation set, we can obtain the combination coefficients $a_{i}^{q}$ and $b_{i}^{q}$ for QP $q$ in depth $i$:
\begin{equation}
\begin{cases}
& a_{i}^{q} = (p_{i}^{q} - p_{i}^{n})/(p_{i}^{m} - p_{i}^{n}) \\
\\
& b_{i}^{q} = (p_{i}^{m} - p_{i}^{q})/(p_{i}^{m} - p_{i}^{n}).
\end{cases}
\end{equation}
Such coefficients are stored as prior information for further usage. We can obtain the prediction vectors $y_{i}^{q}$ of QP $q$ by interpolation as follows:
\begin{equation}
y_{i}^{q} = a_{i}^{q}\cdot y_{i}^{m}+b_{i}^{q}\cdot y_{i}^{n}.
\end{equation}

\section{Fast Intra-Mode Decision}
\subsection{Minimum Number of RDO Candidates}
\begin{figure}[h]
	\vspace{-0.3cm}
	\centerline{\includegraphics[scale=.4]{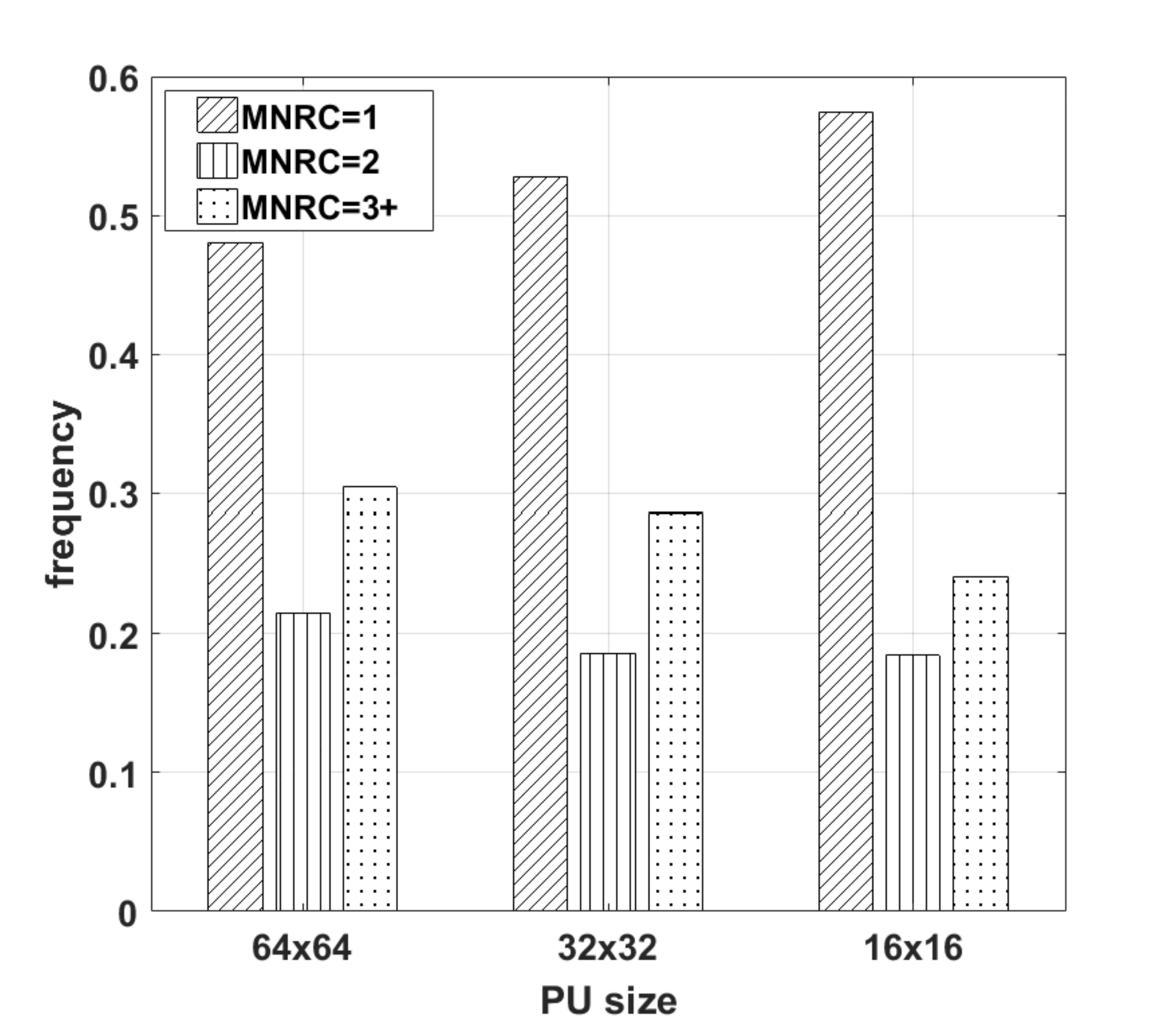}}	
	\caption{The MNRC distribution of PUs in depth 0$\sim$2 for sequence BQMall in class C when QP=32.}
	\label{mnrc}
\end{figure}

For every PU, although the three-step fast intra-mode decision scheme reduces the number of modes for RDO, there is still room for further improvement. The final number of candidate modes (3 for PUs from 64$\times$64 to 16$\times$16 and 8 for 8$\times$8 and 4$\times$4) guarantees coding performance since all these candidate modes need to go through RDO calculation to get the best mode. However, it is unreasonable to indiscriminately perform the same RDO calculation procedure on all PUs in one depth, which results in many redundant computations. In this paper, we define the position (rank) of the best mode in the candidate list generated from the RMD procedure as the minimum number of RDO candidates (MNRC), which can be considered as a flexible threshold to maintain the RD performance. To investigate the characteristics of MNRC, we obtain its distribution from sequence BQMall, which is shown in Fig. \ref{mnrc}. It clearly illustrates that there are still lots of redundancies (for MNRC \textless \ 3) based on the existing three-step scheme.

\begin{figure*}[!h]
	\centerline{\includegraphics[scale=.42]{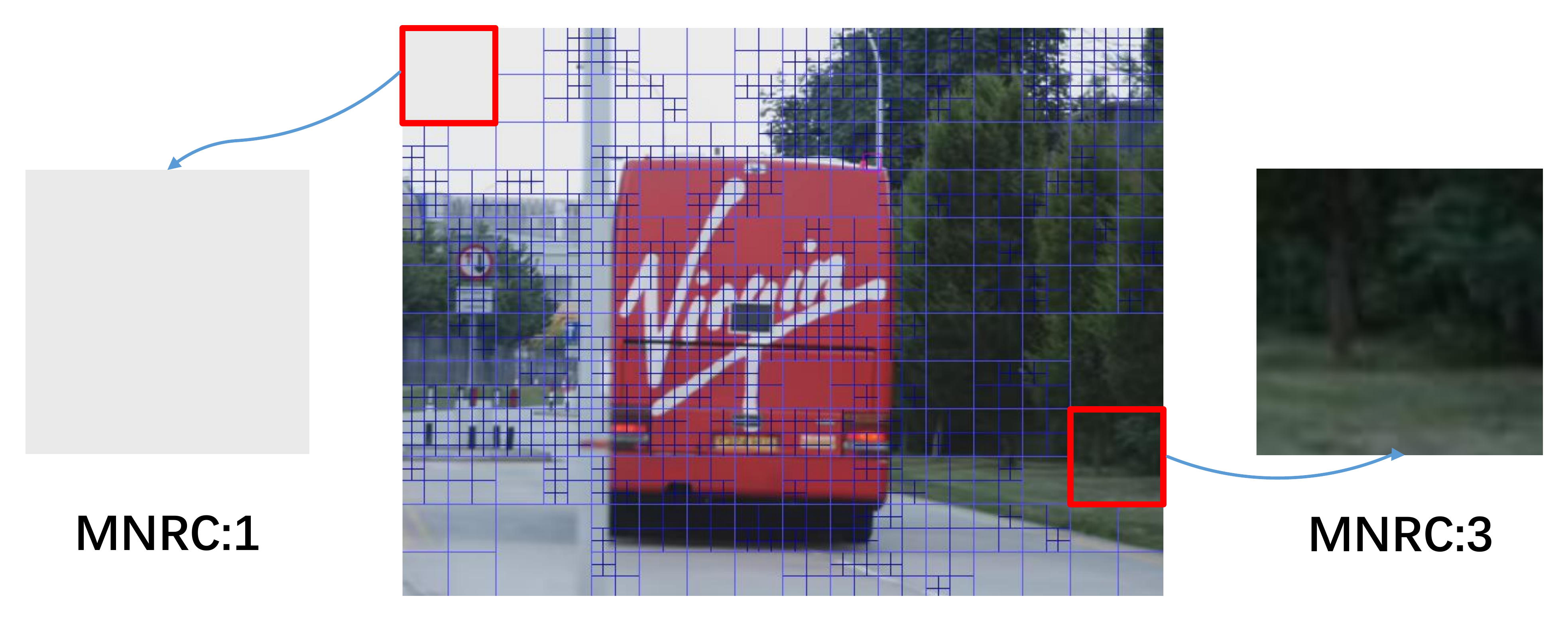}}
	\vspace{-0.3cm}
	\centering
	\caption{A specific case of MNRC. The complex block is likely to be difficult for intra prediction; thus, the MNRC tends to be larger.}
	\label{mnrc-sample}
\end{figure*}

If the MNRC of every PU can be obtained in advance, then we can adaptively set the number of RDO candidate modes for the PU (according to the MNRC); thus, the encoding complexity can be reduced without degrading the RD performance. Thus, the question becomes how to get the MNRC accurately. Fig. \ref{mnrc-sample} gives an example of MNRC corresponding to PUs. We can observe that the complex blocks (difficult for intra prediction) tend to have larger MNRCs, while simple and flat blocks have smaller MNRCs. Therefore, in this paper, we can assume that the MNRC is strongly related to the content of the block. Thus, it is possible and reasonable to predict the MNRC from the block content itself in advance.

\subsection{Expectation Regression Model}
Now, we need to model the relationship between MNRC and the block itself. Similar to the above fast CU/PU size decision, we also employ CNN as the predictor.

\begin{table}[]
	\renewcommand{\arraystretch}{1.5}
	\caption{Designed number of candidate modes for RDO in three categories}
	\centering
	\footnotesize
	\begin{tabular}{cccccc}
		\bottomrule
		& 64x64 & 32x32 & 16x16 & 8x8 & 4x4 \\\midrule
		\multicolumn{1}{c|}{Class 1} & 1     & 1     & 1     & 2   & 2   \\
		\multicolumn{1}{c|}{Class 2} & 2     & 2     & 2     & 5   & 5   \\
		\multicolumn{1}{c|}{Class 3} & 3     & 3     & 3     & 8   & 8  
		\\\bottomrule
	\end{tabular}
\label{tab-num}
\end{table}

For the construction of the model, the intuitive thought is to treat it as a traditional Softmax-based classification problem or to use a single output-node network to regress the MNRC value. However, these methods make the training extremely difficult and lack the ability of fine-grained refinement. Instead, we propose an expectation regression model to solve this problem.

 First, we simplify the alternative actions to 3 in all depths by classifying the 8 RDO candidate modes for PUs 8$\times$8 and 4$\times$4 into 3 categories, as reported in Table \ref{tab-num}, which is aimed at alleviating the training difficulty. Then, CNNs are employed to regress the expectation vector of these three actions in different situations, which can be deemed as a special case of fitting the action-value function (\textit{i.e.}, Q-function \cite{sutton1998reinforcement}) in reinforcement learning. During the test phase, we choose the action with the maximal expectation value. The transformation from the MNRC label to the expectation value is reported in Table \ref{tab-expect} for the following reasons:

\begin{enumerate}
	\item For the correct action, complexity reduction and coding performance achieve the optimal as we expect, so the expectation value is set to $1$. 
	\item If the chosen action is lower than the ground-truth MNRC gear, then it is detrimental to the RD performance since the number of modes to go through RDO calculation is not enough, which should be avoided, so we set the expectation value to 0 in this situation.
	\item For the last case, \textit{i.e.}, the chosen action is higher than the ground-truth MNRC gear, there is no RD performance degradation but some extra complexity, so we set the value to $p$ or $q$, depending on the distance between the chosen action and the ground truth. 
\end{enumerate}

\begin{table}[h]
	\renewcommand{\arraystretch}{1.5}
	\caption{Expectation values according to the MNRC label and corresponding prediction action}
	\centering
	\footnotesize
	\begin{tabular}{cccc}
		\bottomrule
		& Pred 1 & Pred 2 & Pred 3 \\\midrule
		\multicolumn{1}{c|}{Class 1} & 1      & $p$      & $q$      \\
		\multicolumn{1}{c|}{Class 2} & 0      & 1      & $p$      \\
		\multicolumn{1}{c|}{Class 3} & 0      & 0      & 1  
		\\\bottomrule   
	\end{tabular}
	\begin{tablenotes}
		\item \quad \quad \quad \quad  \quad \, 0  $<$ $q$ $<$  $p$ $<$  1
	\end{tablenotes}
	\vspace{-0.cm}
	\label{tab-expect}
\end{table}

By adjusting the parameters of $p$ and $q$, we can obtain models achieving different tradeoffs between complexity reduction and coding performance. Low coding performance degradation can be achieved with relatively low complexity reduction rate, while more complexity can be reduced at the expense of relatively worse coding performance. To validate its effectiveness, we train 2 types of CNNs for fast intra-mode decision, including a conservative setting and an aggressive setting, which focus on the coding performance and the complexity reduction, respectively.

For the CNN structure, we adopt the AK-CNN in the fast CU/PU size decision part. The network is very similar except for the output layer, where there are three output-nodes in this layer (for PU size 4$\times$4, given that the input information is very little, we adopt a shallower and simpler network instead of AK-CNN). 

In our work, CNNs are trained with batches, the size of which is $N$. The expectation value vectors of the ground-truth labels and the prediction outputs are denoted as $y_{i}^{'}$ and $y_{i}$, respectively. The training loss of the fast intra-mode decision is defined as 
\begin{equation}
L_{intra-mode} = \frac{1}{N}\sum_{i=i}^{N}(y_{i}^{'}-y_{i})^{2}.
\end{equation}
During the test phase, the final action $a_{i}^{*}$ is picked as the position of the maximal value in the prediction output vector:
\begin{equation}
a_{i}^{*} = \mathop{\arg\max}_{a_{i}} y_{i}(a_{i}).
\end{equation}

\section{Experimental Results and Analysis}
In this section, we describe the details of our experiments. Part A describes the dataset we established. In parts B and C, we present the configuration of our experiments and the training details of our CNN models. The results and analysis of fast CU/PU size decision and fast intra-mode decision are detailed in part D and part E, respectively. Then, we reveal the result of our overall framework in part F. Next, we analyze the effectiveness of the proposed structure in part G. Finally, the  complexity analysis of our framework is shown in part H.

\subsection{Extended HEVC Intra Coding (EHIC) Dataset}

\begin{table*}[]
	\centering
	\renewcommand{\arraystretch}{1.33}
	\caption{Results for the JCT-VC test set of fast CU/PU size decision}
	\setlength{\tabcolsep}{2.3mm}
	\begin{tabular}{clcccccccccccc}
		\hline
		\multirow{2}{*}{Class} & \multicolumn{1}{c}{\multirow{2}{*}{Sequence}} & \multicolumn{2}{c}{TIP-16 \cite{7547305}}                                                                               & \multicolumn{2}{c}{TCSVT-17 \cite{7457241}}                                                                             & \multicolumn{2}{c}{TIP-18 \cite{8384310}}                                                                               & \multicolumn{2}{c}{LFSD-1}                                                                                 & \multicolumn{2}{c}{LFSD-2}                                                                                 & \multicolumn{2}{c}{LFSD-3}                                                                                 \\ \cline{3-14} 
		& \multicolumn{1}{c}{}                          & \begin{tabular}[c]{@{}c@{}}BD-BR\\ (\%)\end{tabular} & \begin{tabular}[c]{@{}c@{}}$\Delta$T\\ (\%)\end{tabular} & \begin{tabular}[c]{@{}c@{}}BD-BR\\ (\%)\end{tabular} & \begin{tabular}[c]{@{}c@{}}$\Delta$T\\ (\%)\end{tabular} & \begin{tabular}[c]{@{}c@{}}BD-BR\\ (\%)\end{tabular} & \begin{tabular}[c]{@{}c@{}}$\Delta$T\\ (\%)\end{tabular} & \begin{tabular}[c]{@{}c@{}}BD-BR\\ (\%)\end{tabular} & \begin{tabular}[c]{@{}c@{}}$\Delta$T\\ (\%)\end{tabular} & \begin{tabular}[c]{@{}c@{}}BD-BR\\ (\%)\end{tabular} & \begin{tabular}[c]{@{}c@{}}$\Delta$T\\ (\%)\end{tabular} & \begin{tabular}[c]{@{}c@{}}BD-BR\\ (\%)\end{tabular} & \begin{tabular}[c]{@{}c@{}}$\Delta$T\\ (\%)\end{tabular} \\ \hline
		\multirow{2}{*}{A}     & PeopleOnStreet                                & 2.27                                                 & 61.8                                              & 0.63                                                 & 40.9                                              & 2.37                                                 & 61.0                                              & 0.06                                                 & 41.6                                              & 2.10                                                 & 71.6                                              & 3.61                                                 & 77.8                                              \\
		& Traffic                                       & 2.35                                                 & 60.7                                              & 0.61                                                 & 42.4                                              & 2.55                                                 & 70.8                                              & 0.05                                                 & 42.7                                              & 1.95                                                 & 73.0                                              & 3.31                                                 & 79.9                                              \\
		\rowcolor{mygray}
		\multicolumn{2}{c}{Class A Average}                                    & 2.31                                                 & 61.3                                              & 0.62                                                 & 41.7                                              & 2.46                                                 & 65.9                                              & 0.06                                                 & 42.2                                              & 2.03                                                 & 72.3                                              & 3.46                                                 & 78.9                                              \\ \hline
		\multirow{5}{*}{B}     & BasketballDrive                               & 3.49                                                 & 70.9                                              & 0.49                                                 & 54.5                                              & 4.27                                                 & 76.3                                              & 0.09                                                 & 47.8                                              & 2.56                                                 & 76.3                                              & 4.26                                                 & 81.6                                              \\
		& BQTerrace                                     & 2.26                                                 & 62.4                                              & 0.43                                                 & 39.1                                              & 1.84                                                 & 64.7                                              & 0.12                                                 & 43.4                                              & 1.97                                                 & 69.4                                              & 3.36                                                 & 75.9                                              \\
		& Cactus                                        & 2.59                                                 & 60.4                                              & 0.54                                                 & 40.7                                              & 2.27                                                 & 61.0                                              & 0.07                                                 & 42.9                                              & 1.95                                                 & 72.8                                              & 3.33                                                 & 78.8                                              \\
		& Kimono                                        & 2.47                                                 & 62.6                                              & 0.63                                                 & 58.9                                              & 2.59                                                 & 83.5                                              & 0.07                                                 & 54.6                                              & 1.55                                                 & 83.6                                              & 2.30                                                 & 86.1                                              \\
		& ParkScene                                     & 1.86                                                 & 60.3                                              & 0.57                                                 & 45.2                                              & 1.96                                                 & 67.5                                              & 0.03                                                 & 38.2                                              & 1.54                                                 & 73.6                                              & 2.56                                                 & 79.7                                              \\
		\rowcolor{mygray}
		\multicolumn{2}{c}{Class B Average}                                    & 2.53                                                 & 63.3                                              & 0.53                                                 & 47.7                                              & 2.59                                                 & 70.6                                              & 0.08                                                 & 45.4                                              & 1.91                                                 & 75.1                                              & 2.89                                                 & 80.4                                              \\ \hline
		\multirow{4}{*}{C}     & BasketballDrill                               & 4.26                                                 & 56.9                                              & 0.59                                                 & 36.6                                              & 2.86                                                 & 53.0                                              & 0.29                                                 & 41.2                                              & 2.97                                                 & 67.3                                              & 4.82                                                 & 73.5                                              \\
		& BQMall                                        & 2.93                                                 & 57.8                                              & 0.33                                                 & 36.5                                              & 2.09                                                 & 58.4                                              & 0.09                                                 & 46.9                                              & 1.72                                                 & 70.7                                              & 2.96                                                 & 76.1                                              \\
		& PartyScene                                    & 2.19                                                 & 51.1                                              & 0.40                                                 & 28.1                                              & 0.66                                                 & 44.5                                              & 0.04                                                 & 42.5                                              & 0.82                                                 & 63.3                                              & 1.52                                                 & 68.9                                              \\
		& RaceHorses                                    & 2.08                                                 & 56.2                                              & 0.53                                                 & 42.7                                              & 1.97                                                 & 57.1                                              & 0.13                                                 & 44.0                                              & 1.95                                                 & 70.7                                              & 3.13                                                 & 76.3                                              \\
		\rowcolor{mygray}
		\multicolumn{2}{c}{Class C Average}                                    & 2.87                                                 & 55.5                                              & 0.46                                                 & 36.0                                              & 1.90                                                 & 53.3                                              & 0.14                                                 & 43.7                                              & 1.87                                                 & 68.0                                              & 3.11                                                 & 73.7                                              \\ \hline
		\multirow{4}{*}{D}     & BasketballPass                                & 2.89                                                 & 59.3                                              & 0.43                                                 & 37.0                                              & 1.84                                                 & 56.4                                              & 0.06                                                 & 43.9                                              & 1.63                                                 & 70.3                                              & 2.73                                                 & 76.3                                              \\
		& BlowingBubbles                                & 2.54                                                 & 53.6                                              & 0.27                                                 & 28.7                                              & 0.62                                                 & 40.5                                              & 0.04                                                 & 40.3                                              & 0.95                                                 & 61.5                                              & 1.73                                                 & 67.8                                              \\
		& BQSquare                                      & 1.48                                                 & 52.6                                              & 0.32                                                 & 33.2                                              & 0.91                                                 & 45.8                                              & 0.08                                                 & 47.6                                              & 0.91                                                 & 64.3                                              & 1.56                                                 & 69.1                                              \\
		& RaceHorese                                    & 2.43                                                 & 53.6                                              & 0.29                                                 & 34.3                                              & 1.32                                                 & 55.8                                              & 0.07                                                 & 41.7                                              & 1.47                                                 & 67.7                                              & 2.53                                                 & 73.8                                              \\
		\rowcolor{mygray}
		\multicolumn{2}{c}{Class D Average}                                    & 2.34                                                 & 54.8                                              & 0.33                                                 & 33.3                                              & 1.17                                                 & 49.6                                              & 0.06                                                 & 43.4                                              & 1.24                                                 & 66.0                                              & 2.14                                                 & 71.8                                              \\ \hline
		\multirow{3}{*}{E}     & FourPeople                                    & 3.05                                                 & 60.0                                                & 0.70                                                 & 46.2                                              & 3.11                                                 & 71.3                                              & 0.08                                                 & 50.6                                              & 2.39                                                 & 75.1                                              & 4.03                                                 & 80.0                                              \\
		& Johnny                                        & 4.42                                                 & 72.2                                              & 1.01                                                 & 56.8                                              & 3.82                                                 & 70.7                                              & 0.17                                                 & 59.6                                              & 2.73                                                 & 81.1                                              & 4.19                                                 & 84.0                                              \\
		& KristenAndSara                                & 3.12                                                 & 69.3                                              & 0.70                                                 & 53.4                                              & 3.46                                                 & 74.8                                              & 0.11                                                 & 56.8                                              & 2.28                                                 & 79.1                                              & 3.78                                                 & 82.6                                              \\
		\rowcolor{mygray}
		\multicolumn{2}{c}{Class E Average}                                    & 3.53                                                 & 67.2                                              & 0.80                                                 & 52.1                                              & 3.46                                                 & 72.3                                              & 0.12                                                 & 55.7                                              & 2.47                                                 & 78.4                                              & 4.00                                                 & 82.2                                              \\ \bottomrule[1.5pt]
		\rowcolor{mygray}
		\multicolumn{2}{c}{\textbf{Average}}                                   & \textbf{2.54}                                        & \textbf{60.1}                                     & \textbf{0.53}                                        & \textbf{42.0}                                     & \textbf{2.25}                                        & \textbf{61.8}                                     & \textbf{0.09}                                        & \textbf{45.9}                                     & \textbf{1.86}                                        & \textbf{71.7}                                     & \textbf{3.10}                                        & \textbf{77.1}                                     \\ \bottomrule[1.5pt]
	\end{tabular}
\begin{tablenotes}
	\item \ \ \ \ LFSD-1, LFSD-3 indicate the leftmost and rightmost points, i.e., the LR and HR modes, while LFSD-2 is for the optimal tradeoff (OT) point.
\end{tablenotes}
\label{tab-size}
\end{table*}

As CNN is employed as the classifier in our work, a large amount of training data is required. Although a CU partition dataset CPH \cite{8384310} already exists, it does not provide the label for PU partition and intra-mode decision. Thus, we build a complete dataset for extended HEVC intra coding, namely, the EHIC dataset, including fast CU/PU size decision and intra-mode decision. Both high-definition and low-definition raw images are collected from the Uncompressed Color Image Database (UCID) \cite{schaefer2003ucid}, Raw Images Dataset (RAISE)\cite{dang2015raise} and DIVerse 2K Resolution Image Dataset (DIV2K) \cite{agustsson2017ntire}, corresponding to the resolution range of the JCT-VC standard test set \cite{bossen2013test}. These images are randomly divided into a training set (90\%) and validation set (10\%). Then, all the images are encoded by the HEVC reference software HM 16.9, while four QPs in \{22, 27, 32, 37\} are applied for encoding. During encoding, we collect the ranking position of the final intra-prediction mode in the candidate list generated from the RMD procedure for every PU as the MNRC label. After encoding, we collect labels indicating the splitting or nonsplitting operations of all CUs for the training of size decision. Next, the CU/PUs cropped from the images are combined with the corresponding labels as the samples of our dataset.

In addition, the RD-cost of each block is also collected during encoding, which can be used for further optimization of the training process for fast CU/PU size decision. Generally, the coding performance degradation due to misclassification of different blocks is also different. Therefore, it is not reasonable to consider the splitting decision as an indiscriminate classification problem. We define the coding performance loss $Loss_{RD}$ as follows: 
\begin{equation}
\begin{cases}
& Loss_{RD}=\frac{|J_{before} - J_{after}|}{J_{before} + J_{after}} \\
\\
& J=SSE+\lambda\cdot R_{total}\
\end{cases}
\end{equation}
Here, $J_{before}$ and $J_{after}$ are the RD-cost values of the block before and after splitting processing, respectively.
Based on such evaluating indicator, we introduce a new format of loss function to optimize the training for fast CU/PU size decision. Similar to \textit{focal loss} \cite{lin2018focal}, the loss function of block with small $Loss_{RD}$ is allocated with a small weight $w$. The optimized training loss function is designed as follows: 
\begin{equation}
Loss_{size} = 
\begin{cases}
& -w\sum\limits_{i}y_{i}^{'}log(y_{i})\qquad if\ Loss_{RD}<th_{RD}\\
\\
& -\sum\limits_{i}y_{i}^{'}log(y_{i}) \qquad \ \ \ \ otherwise.
\end{cases}
\end{equation}
Here, $th_{RD}$ is the preset threshold to modify the loss for RD performance, and $y_{i}^{'}$ and $y_{i}$ are the ground-truth labels and the prediction outputs, respectively. 

As a consequence, AK-CNNs will pay more attention to the blocks with large $Loss_{RD}$. The intuition of this design is that the fast decision for the blocks with small $Loss_{RD}$ is of little importance, both splitting or nonsplitting will not result in too much coding performance degradation. This method can bring a better performance.

\subsection{Configuration of Experiments}
In our experiments, all the schemes are implemented into the HEVC reference software HM 16.9, including both the fast CU/PU size decision and fast mode decision. In HM 16.9, the all-intra main configuration is used with the default configuration file \textit{encoder\_intra\_main.cfg} \cite{bossen2013test}, and four QP values in \{22,27,32,37\} are chosen to compress the frames. Our experiments are tested on 18 video sequences of the JCT-VC standard test set.

In our experiments, the coding performance is measured by the Bj$\phi$ntegaard delta bit-rate (BD-BR) \cite{bjontegaard2001calculation} compared with the original HM 16.9, and the encoding time-saving rate $\Delta T$ is calculated as
\begin{equation}
\Delta T = \frac{T_{HM}-T_{test}}{T_{HM}}
\end{equation}
where $T_{test}$ is the encoding time of the proposed method in the test mode and $T_{HM}$ is the encoding time of the anchor of HM 16.9. 

All experiments are conducted on a computer with
an Intel (R) Core (TM) i7-7700K CPU and 16 GB
RAM. An NVIDIA GeForce GTX 1080Ti GPU was used to accelerate the training process, but it was disabled when testing the HEVC complexity reduction. The \textit{TensorFlow} \cite{abadi2016tensorflow} platform is used for our CNN models.

\subsection{CNN Training Details}
Here, we present the details of the CNN training. All the CNN models of our work are
trained from the training set of EHIC dataset, while the hyper-parameters are fine-tuned on the validation set. The convolutional layers and fully connected layers are all initialized from Gaussian distributions. For training these models, we use the Adam optimizer \cite{kingma2014adam} with an initial learning rate of
5$\times$$10^{-3}$ for 150 epochs. The learning rate is divided by a factor of 10 every 50 epochs. Dropout \cite{srivastava2014dropout} is added after all hidden fully connected layers to avoid overfitting, and the drop ratio is set to 0.5. The batch size is 64 for the models of large CU/PU (depth = 0, 1) and 256 for the models of small CU/PU (depth = 2, 3, 4). 
\begin{figure}[h]
	\vspace{-0.2cm}
	\centerline{\includegraphics[scale=.35]{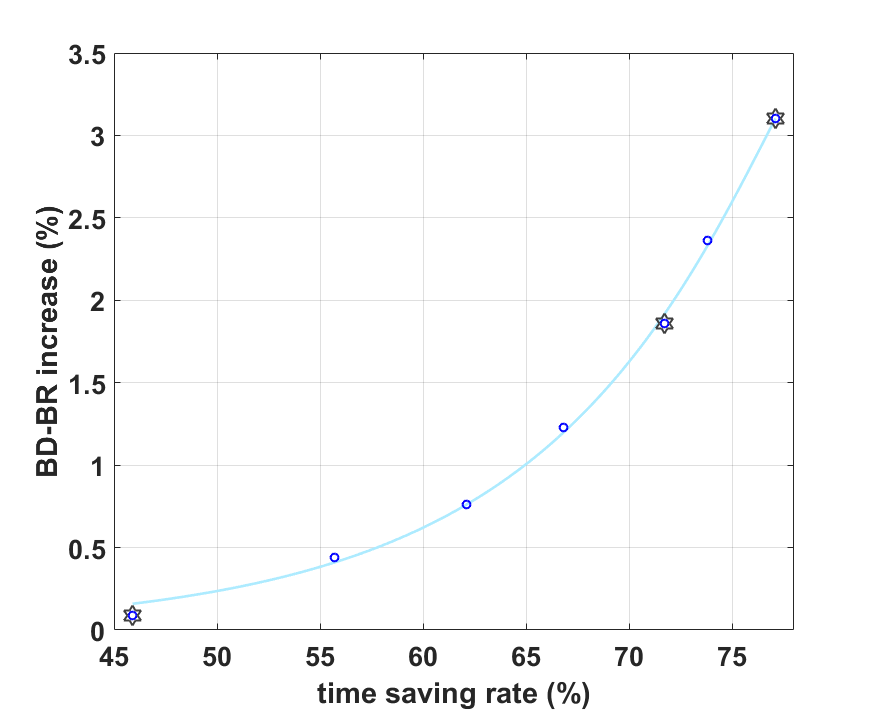}}
	\caption{Configurable tradeoffs arise from EOTD. Hexagon points will be used for further comparison.}
	\vspace{-0.4cm}
	\label{config}
\end{figure}
\begin{table}[!h]
	\renewcommand{\arraystretch}{1.12}
	\caption{Ratio of fast skip in every depth for three modes}
	\centering
	\setlength{\tabcolsep}{2.2mm}
	\begin{tabular}{lllccc}\toprule
		&                          &         & RD Check   & Early Term   & Early Split   \\\midrule
		\multirow{20}{*}{LR} & \multirow{4}{*}{Class A} & Depth 0 & 17.2\% & 0.0\%  & 82.8\% \\
		&                          & Depth 1 & 55.2\% & 3.1\%  & 41.7\% \\
		&                          & Depth 2 & 76.4\% & 13.2\% & 10.4\% \\
		&                          & Depth 3 & 69.1\% & 29.0\% & 1.9\%  \\ \cline{2-6} 
		& \multirow{4}{*}{Class B} & Depth 0 & 46.8\% & 0.2\%  & 53.0\% \\
		&                          & Depth 1 & 61.6\% & 8.8\%  & 29.6\% \\
		&                          & Depth 2 & 70.9\% & 16.3\% & 12.8\% \\
		&                          & Depth 3 & 75.1\% & 22.4\% & 2.5\%  \\ \cline{2-6} 
		& \multirow{4}{*}{Class C} & Depth 0 & 17.0\% & 0.0\%  & 83.0\% \\
		&                          & Depth 1 & 29.3\% & 1.4\%  & 69.3\% \\
		&                          & Depth 2 & 60.0\% & 8.1\%  & 31.9\% \\
		&                          & Depth 3 & 73.8\% & 19.6\% & 6.6\%  \\ \cline{2-6} 
		& \multirow{4}{*}{Class D} & Depth 0 & 11.8\% & 0.0\%  & 88.2\% \\
		&                          & Depth 1 & 27.7\% & 0.8\%  & 71.5\% \\
		&                          & Depth 2 & 52.7\% & 6.2\%  & 41.1\% \\
		&                          & Depth 3 & 71.3\% & 17.0\% & 11.7\% \\ \cline{2-6} 
		& \multirow{4}{*}{Class E} & Depth 0 & 42.5\% & 0.6\%  & 56.9\% \\
		&                          & Depth 1 & 54.2\% & 12.4\% & 33.4\% \\
		&                          & Depth 2 & 63.5\% & 25.0\% & 11.5\% \\
		&                          & Depth 3 & 63.4\% & 33.3\% & 3.3\% \\\midrule
		\multirow{20}{*}{OT} & \multirow{4}{*}{Class A} & Depth 0 & 0.0\%  & 2.4\%  & 97.6\% \\
		&                          & Depth 1 & 5.4\%  & 21.9\% & 72.7\% \\
		&                          & Depth 2 & 18.4\% & 39.7\% & 41.9\% \\
		&                          & Depth 3 & 23.8\% & 59.9\% & 16.3\% \\ \cline{2-6} 
		& \multirow{4}{*}{Class B} & Depth 0 & 0.0\%  & 6.1\%  & 93.9\% \\
		&                          & Depth 1 & 6.7\%  & 40.5\% & 52.8\% \\
		&                          & Depth 2 & 17.9\% & 41.3\% & 40.8\% \\
		&                          & Depth 3 & 29.9\% & 51.3\% & 18.8\% \\ \cline{2-6} 
		& \multirow{4}{*}{Class C} & Depth 0 & 0.0\%  & 0.6\%  & 99.4\% \\
		&                          & Depth 1 & 2.9\%  & 8.6\%  & 88.5\% \\
		&                          & Depth 2 & 13.0\% & 22.6\% & 64.4\% \\
		&                          & Depth 3 & 26.9\% & 41.8\% & 31.3\% \\ \cline{2-6} 
		& \multirow{4}{*}{Class D} & Depth 0 & 0.0\%  & 0.1\%  & 99.9\% \\
		&                          & Depth 1 & 3.0\%  & 8.2\%  & 88.8\% \\
		&                          & Depth 2 & 11.2\% & 19.0\% & 69.8\% \\
		&                          & Depth 3 & 24.6\% & 36.5\% & 38.9\% \\ \cline{2-6} 
		& \multirow{4}{*}{Class E} & Depth 0 & 0.0\%  & 18.7\% & 81.3\% \\
		&                          & Depth 1 & 5.1\%  & 39.9\%& 55.0\% \\
		&                          & Depth 2 & 16.4\% & 47.5\% & 36.1\% \\
		&                          & Depth 3 & 22.6\% & 62.1\% & 15.3\% \\\midrule
		\multirow{20}{*}{HR} & \multirow{4}{*}{Class A} & Depth 0 & 0.0\%  & 2.4\% & 97.6\% \\
		&                          & Depth 1 & 0.0\%  & 24.4\% & 75.6\% \\
		&                          & Depth 2 & 0.9\%  & 47.1\% & 52.0\% \\
		&                          & Depth 3 & 10.1\% & 67.8\% & 22.1\% \\ \cline{2-6} 
		& \multirow{4}{*}{Class B} & Depth 0 & 0.0\%  & 6.1\%  & 93.9\% \\
		&                          & Depth 1 & 0.0\%  & 43.9\% & 56.1\% \\
		&                          & Depth 2 & 0.8\%  & 48.3\% & 50.9\% \\
		&                          & Depth 3 & 13.7\% & 60.9\% & 25.4\% \\ \cline{2-6} 
		& \multirow{4}{*}{Class C} & Depth 0 & 0.0\%  & 0.6\%  & 99.4\% \\
		&                          & Depth 1 & 0.0\%  & 10.0\% & 90.0\% \\
		&                          & Depth 2 & 0.7\%  & 27.6\% & 71.7\% \\
		&                          & Depth 3 & 11.1\% & 49.5\% & 39.3\% \\ \cline{2-6} 
		& \multirow{4}{*}{Class D} & Depth 0 & 0.0\%  & 0.1\%  & 99.9\% \\
		&                          & Depth 1 & 0.0\%  & 9.7\%  & 90.3\% \\
		&                          & Depth 2 & 0.6\%  & 23.0\% & 76.4\% \\
		&                          & Depth 3 & 9.9\%  & 43.5\% & 46.6\% \\ \cline{2-6} 
		& \multirow{4}{*}{Class E} & Depth 0 & 0.0\%  & 18.7\% & 81.3\% \\
		&                          & Depth 1 & 0.0\%  & 42.2\% & 57.8\% \\
		&                          & Depth 2 & 0.8\%  & 54.2\% & 45.0\% \\
		&                          & Depth 3 & 9.8\%  & 69.7\% & 20.5\% \\\bottomrule
	\end{tabular}
	\begin{tablenotes}
		\item Early Termination is abbreviated as Early Term in this table.
	\end{tablenotes}
\vspace{-0.6cm}
\label{tab-ratio}
\end{table}

\subsection{Performance Evaluation of CU/PU size decision}
\textbf{Scalable complexity reduction.}
First, we validate the scalable complexity reduction ability of our approach. As we use the confidence thresholds in four depths, different tradeoffs between RD performance and complexity reduction are achieved. Then, we obtain the best combinations of the threshold values (\textit{Pareto optimal points}) from the evolutionary algorithm. According to the different combinations of thresholds, we enable our framework with fine-grained scalability in complexity reduction. Fig. \ref{config} shows the results of RD performance degradation and the complexity reduction, which are averaged at four QPs in \{22,27,32,37\} over the 18 JCT-VC test sequences. From Fig. \ref{config}, we can observe that the tradeoff between the RD performance and complexity reduction is well kept: we can achieve almost perfect RD performance maintenance with low complexity reduction or reduce a large percentage of encoding complexity with a relatively larger BD-BR increase. This scalability enhances the usability of our approach, and based on this, we can achieve different levels of coding complexity reduction according to different requirements of RD performance degradation.

\begin{figure*}[!h]
	\centerline{\includegraphics[scale=.27]{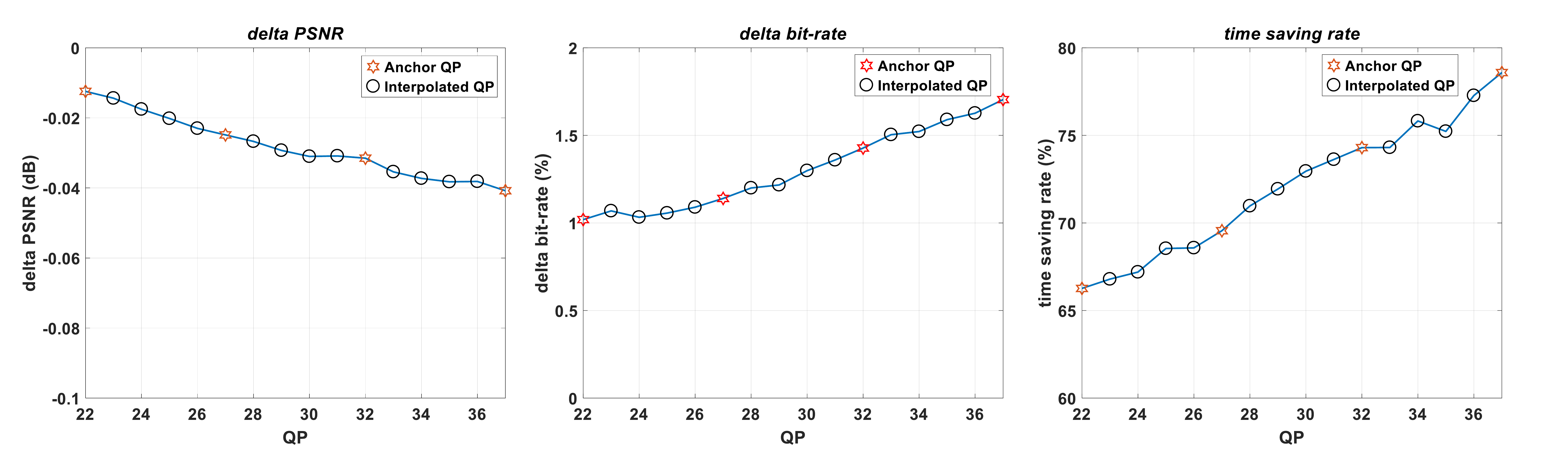}}
	\caption{Generalization capability among a large range of QPs.}
	\vspace{-0.2cm}
	\label{qp}
\end{figure*}

\textbf{Evaluation in terms of complexity reduction.}
Then, we evaluate the performance of our learned fast size decision (LFSD) in terms of complexity reduction. We pick three points in Fig. \ref{config} (the hexagon marks) to represent our scalable fast approach. The leftmost and rightmost points are chosen, which can achieve extreme RD performance maintaining and complexity reduction, namely, the low RD-degradation (LR) and the high RD-degradation (HR) modes, respectively. In addition, we also pick a middle point to represent the balanced tradeoff between RD performance and complexity reduction, which is called the optimal tradeoff (OT) mode. Then, we compare them with three state-of-the-art results \cite{7547305,7457241} and \cite{8384310}, as reported in Table \ref{tab-size}. In the LR mode, the fast algorithm is executed only if the AK-CNNs are exceedingly confident with the decisions, so it reduces the encoding complexity merely by 45.9\%, which is still better than \cite{7457241} ($\Delta T = 42.0\%$). As for the OT and HR modes, the ratio of fast decision is much higher. Therefore, these two schemes reduce the encoding complexity by 71.7\% and 77.1\%, respectively. These performances significantly outperform \cite{7547305} ($\Delta T = 60.1\%$) and \cite{8384310} ($\Delta T = 61.8\%$), especially the HR mode, saving an additional 17.0\% and 15.3\% of the encoding time. Since our approach implements the complete fast CTU partition operation, including fast PU size decision, it can achieve lower complexity of HEVC intra coding than others.

Here, we also analyze where and how our approach reduces the encoding complexity. We count and then calculate the ratio of blocks on which we implement fast skip (including early split and early termination) and RD check in four depths for the above three modes during encoding, as reported in Table \ref{tab-ratio}, the results in which are averaged over four QPs in \{22,27,32,37\}. For the LR mode, since our algorithm is very conservative, only 48.1\% RD checks are omitted on average. Most of the fast skips are distributed at depth 0, as over 80\% of 64$\times$64 CUs are early split safely except for class B and class E. For the sequences in these two classes, there are plenty of flat background areas, so the possibility of large CU size is quite high. As a result, RD checks are needed for such blocks (although the definition of sequences in class A is high, there are very few smooth large blocks). As for the aggressive HR mode, almost all the RD checks are skipped, with only some for the small 8$\times$8 CUs remaining due to its difficultly of prediction. Consequently, the HR mode can reduce the encoding complexity dramatically. To achieve the balance, the OT mode skips plenty of the RD checks at the high level (for large CU sizes), while it implements some necessary RD checks for small blocks. Thus, it can greatly reduce the complexity, without incurring too much RD performance degradation. Obviously, our algorithm can intelligently execute the check or skip actions according to the characteristics of the content.

\textbf{Evaluation on RD performance.}
Next, we compare the RD performance with other three approaches in terms of BD-BR. From Table \ref{tab-size}, we can observe that our approach of LR mode incurs a mere 0.09\% BD-BR increase, almost the same as the anchor of HM 16.9, much better than \cite{7457241} (0.53\% increase). For the OT mode, the BD-BR increase is on average 1.86\%,  which significantly outperforms \cite{7547305} (2.54\% increase) and \cite{8384310} (2.25\% increase). This promising result owes to the high accuracies of the AK-CNNs and the confidence threshold control. As for the HR mode, the BD-BR increase reaches 3.10\%. Due to the requirement of high complexity reduction, many nonconfident decisions are made to accelerate the encoding; thus, the RD performance degradation is relatively high.

 \begin{table*}[!!h]
 	\centering
 	\renewcommand{\arraystretch}{1.35}
 	\caption{Results for the JCT-VC test set of fast intra-mode decision and overall performance}
 	\setlength{\tabcolsep}{2.3mm}
 	\begin{tabular}{clcccccccccc}
 		\hline
 		\multirow{2}{*}{Class} & \multicolumn{1}{c}{\multirow{2}{*}{Sequence}} & \multicolumn{2}{c}{TCSVT-17 \cite{7457241}}                                                                             & \multicolumn{2}{c}{TIP-18 \cite{8412615}}                                                                               & \multicolumn{2}{c}{LFMD*}                                                                                & \multicolumn{2}{c}{LFMD}                                                                                 & \multicolumn{2}{c}{LFHI}                                                                              \\ \cline{3-12} 
 		& \multicolumn{1}{c}{}                          & \begin{tabular}[c]{@{}c@{}}BD-BR\\ (\%)\end{tabular} & \begin{tabular}[c]{@{}c@{}}$\Delta$T\\ (\%)\end{tabular} & \begin{tabular}[c]{@{}c@{}}BD-BR\\ (\%)\end{tabular} & \begin{tabular}[c]{@{}c@{}}$\Delta$T\\ (\%)\end{tabular} & \begin{tabular}[c]{@{}c@{}}BD-BR\\ (\%)\end{tabular} & \begin{tabular}[c]{@{}c@{}}$\Delta$T\\ (\%)\end{tabular} & \begin{tabular}[c]{@{}c@{}}BD-BR\\ (\%)\end{tabular} & \begin{tabular}[c]{@{}c@{}}$\Delta$T\\ (\%)\end{tabular} & \begin{tabular}[c]{@{}c@{}}BD-BR\\ (\%)\end{tabular} & \begin{tabular}[c]{@{}c@{}}$\Delta$T\\ (\%)\end{tabular} \\ \hline
 		\multirow{2}{*}{A}     & PeopleOnStreet                                & 0.29                                                 & 16.5                                              & 0.50                                                 & 19.3                                              & 0.38                                                 & 25.1                                              & 0.19                                                 & 20.9                                              & 2.35                                                 & 75.5                                              \\
 		& Traffic                                       & 0.15                                                 & 17.6                                              & 0.50                                                 & 16.5                                              & 0.37                                                 & 24.6                                              & 0.19                                                 & 21.2                                              & 2.20                                                 & 76.8                                              \\
 		\rowcolor{mygray}
 		\multicolumn{2}{c}{Class A Average}                                    & 0.22                                                 & 17.1                                              & 0.50                                                 & 17.9                                              & 0.38                                                 & 24.9                                              & 0.19                                                 & 21.1                                              & 2.28                                                 & 76.2                                              \\ \hline
 		\multirow{5}{*}{B}     & BasketballDrive                               & 0.12                                                 & 16.3                                              & 0.50                                                 & 17.3                                              & 0.36                                                 & 26.2                                              & 0.19                                                 & 22.9                                              & 2.78                                                 & 79.0                                              \\
 		& BQTerrace                                     & 0.15                                                 & 17.4                                              & 0.40                                                 & 16.8                                              & 0.17                                                 & 26.7                                              & 0.08                                                 & 21.7                                              & 2.08                                                 & 73.3                                              \\
 		& Cactus                                        & 0.13                                                 & 12.3                                              & 0.50                                                 & 18.5                                              & 0.32                                                 & 25.1                                              & 0.16                                                 & 21.2                                              & 2.14                                                 & 75.7                                              \\
 		& Kimono                                        & -0.01                                                & 15.6                                              & 0.10                                                 & 19.8                                              & 0.21                                                 & 26.4                                              & 0.12                                                 & 23.2                                              & 1.67                                                 & 84.5                                              \\
 		& ParkScene                                     & 0.09                                                 & 12.7                                              & 0.30                                                 & 18.3                                              & 0.25                                                 & 24.1                                              & 0.13                                                 & 20.4                                              & 1.70                                                 & 75.9                                              \\
 		\rowcolor{mygray}
 		\multicolumn{2}{c}{Class B Average}                                    & 0.10                                                 & 14.9                                              & 0.36                                                 & 18.1                                              & 0.26                                                 & 25.7                                              & 0.14                                                 & 21.9                                              & 2.07                                                 & 77.7                                              \\ \hline
 		\multirow{4}{*}{C}     & BasketballDrill                               & 0.36                                                 & 11.0                                              & 0.50                                                 & 16.1                                              & 0.32                                                 & 25.6                                              & 0.16                                                 & 22.9                                              & 3.18                                                 & 71.6                                              \\
 		& BQMall                                        & 0.23                                                 & 17.5                                              & 0.60                                                 & 16.3                                              & 0.44                                                 & 25.8                                              & 0.22                                                 & 23.4                                              & 1.99                                                 & 74.8                                              \\
 		& PartyScene                                    & 0.23                                                 & 12.3                                              & 0.80                                                 & 17.3                                              & 0.42                                                 & 24.3                                              & 0.20                                                 & 20.6                                              & 1.07                                                 & 68.5                                              \\
 		& RaceHorses                                    & 0.10                                                 & 15.0                                              & 0.40                                                 & 18.0                                              & 0.26                                                 & 24.0                                              & 0.12                                                 & 21.3                                              & 2.16                                                 & 73.0                                              \\
 		\rowcolor{mygray}
 		\multicolumn{2}{c}{Class C Average}                                    & 0.23                                                 & 14.0                                              & 0.58                                                 & 16.9                                              & 0.36                                                 & 24.9                                              & 0.18                                                 & 22.1                                              & 2.10                                                 & 72.0                                              \\ \hline
 		\multirow{4}{*}{D}     & BasketballPass                                & 0.36                                                 & 18.7                                              & 0.60                                                 & 15.2                                              & 0.46                                                 & 24.8                                              & 0.23                                                 & 21.3                                              & 1.90                                                 & 74.7                                              \\
 		& BlowingBubbles                                & 0.21                                                 & 13.7                                              & 0.70                                                 & 19.5                                              & 0.41                                                 & 22.4                                              & 0.19                                                 & 19.0                                              & 1.17                                                 & 66.6                                              \\
 		& BQSquare                                      & 0.52                                                 & 12.6                                              & 0.80                                                 & 18.9                                              & 0.45                                                 & 23.5                                              & 0.21                                                 & 20.5                                              & 1.20                                                 & 69.4                                              \\
 		& RaceHorese                                    & 0.37                                                 & 13.7                                              & 0.60                                                 & 18.5                                              & 0.41                                                 & 22.4                                              & 0.19                                                 & 19.7                                              & 1.69                                                 & 71.1                                              \\
 		\rowcolor{mygray}
 		\multicolumn{2}{c}{Class D Average}                                    & 0.37                                                 & 14.7                                              & 0.68                                                 & 18.0                                              & 0.43                                                 & 23.3                                              & 0.21                                                 & 20.1                                              & 1.49                                                 & 70.5                                              \\ \hline
 		\multirow{3}{*}{E}     & FourPeople                                    & 0.11                                                 & 22.6                                              & 0.50                                                 & 17.1                                              & 0.43                                                 & 27.1                                              & 0.21                                                 & 25.2                                              & 2.65                                                 & 78.2                                              \\
 		& Johnny                                        & 0.28                                                 & 21.6                                              & 0.50                                                 & 17.5                                              & 0.46                                                 & 26.5                                              & 0.26                                                 & 25.1                                              & 3.05                                                 & 83.0                                              \\
 		& KristenAndSara                                & 0.17                                                 & 23.5                                              & 0.50                                                 & 17.4                                              & 0.44                                                 & 25.8                                              & 0.25                                                 & 24.7                                              & 2.58                                                 & 81.6                                              \\
 		\rowcolor{mygray}
 		\multicolumn{2}{c}{Class E Average}                                    & 0.19                                                 & 22.6                                              & 0.50                                                 & 17.3                                              & 0.44                                                 & 26.5                                              & 0.24                                                 & 25.0                                              & 2.76                                                 & 80.9                                              \\ \bottomrule[1.5pt]
 		\rowcolor{mygray}
 		\multicolumn{2}{c}{\textbf{Average}}                                   & \textbf{0.21}                                        & \textbf{16.1}                                     & \textbf{0.52}                                        & \textbf{17.7}                                     & \textbf{0.36}                                        & \textbf{25.0}                                     & \textbf{0.18}                                        & \textbf{22.0}                                     & \textbf{2.09}                                        & \textbf{75.2}                                     \\ \bottomrule[1.5pt]
 	\end{tabular}
 	\begin{tablenotes}
 		\item LFMD* indicates our aggressive scheme, while LFMD is for the conservative scheme. LFHI is our overall framework, which is the assembled system of LFSD and LFMD.
 	\end{tablenotes}
 	\vspace{-0.3cm}
 	\label{tab-mode}
 \end{table*}
 
\textbf{Generalization capability at different QPs.}
In addition to four QPs evaluated above (QP = 22, 27, 32, 37), we further test our approach for reducing complexity of intra-mode HEVC at other QPs in [22, 37]. To verify the generalization capability of our approach, we use the proposed interpolation-based approach. Fig. \ref{qp} illustrates the bit-rate difference, PSNR loss and encoding time reduction of our approach at different QPs. Note that the results are averaged over the 18 JCT-VC test video sequences. In this figure, the hexagon marks denote the test results at four anchor QPs, whereas the circle marks represent the test results at 12 interpolated QPs. We can see that the performance of the interpolated QPs is very stable. The three curves maintain superb monotonicity and minuscule fluctuate. It is difficult to distinguish the interpolated QPs. This performance is much better than the work that uses one large CNN targeting all QPs, such as \cite{8384310}. As for the time of feed-forward of CNNs, although we have to run the CNN twice for the interpolated QPs, our work still performs well in terms of complexity reduction due to the low complexity of our extremely light network structure (the inference time accounts for less than 1\% of the total encoding time of the anchor of HM 16.9). Obviously, our special interpolation-based design can better satisfy the requirements of different QPs, which is more effective.

\begin{figure}[!!h]
 	\vspace{-0.4cm}
 	\centerline{\includegraphics[scale=.2]{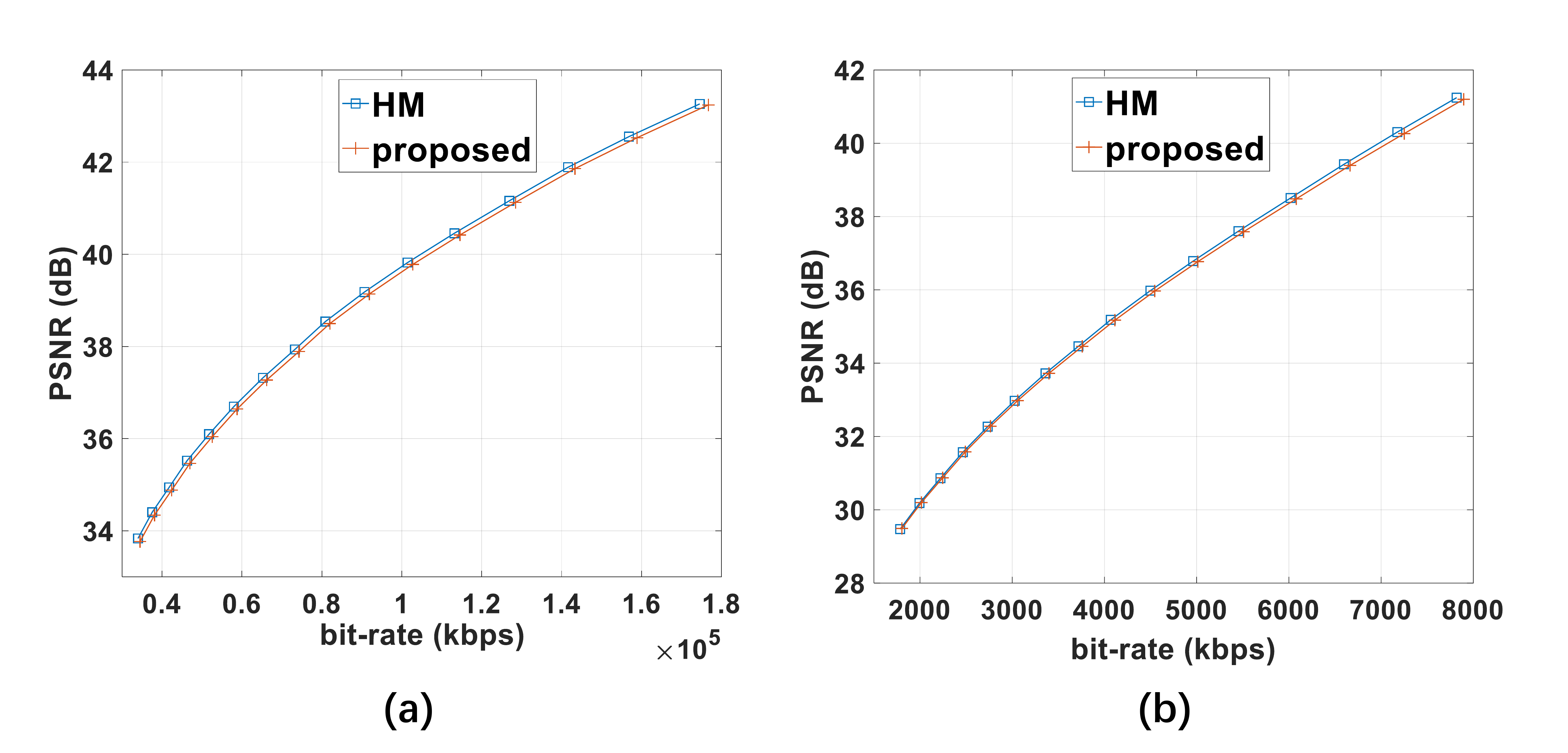}}
 	\caption{RD curve comparison. (a) PeopleOnStreet in Class A. (b) BQMall in Class D.}
 	\vspace{-0.2cm}
 	\label{curve}
 \end{figure}

Furthermore, we provide the RD curves of our approach and the anchor of HM 16.9 for two JCT-VC sequences, including the high-definition sequence, PeopleOnStreet in Class A, and the low-definition sequence, BQMall in Class D. Fig. \ref{curve} shows a comparison of the different curves. From this figure, we can observe that the difference of RD performance between the anchor of HM 16.9 and our approach is very small for all bit-rate points. It shows that our approach can adapt to the different bit-rate points admirably, for both high-definition and low-definition sequences.

\subsection{Performance Evaluation of Mode Decision}
\begin{figure*}[!!h]
 	\centerline{\includegraphics[scale=.16]{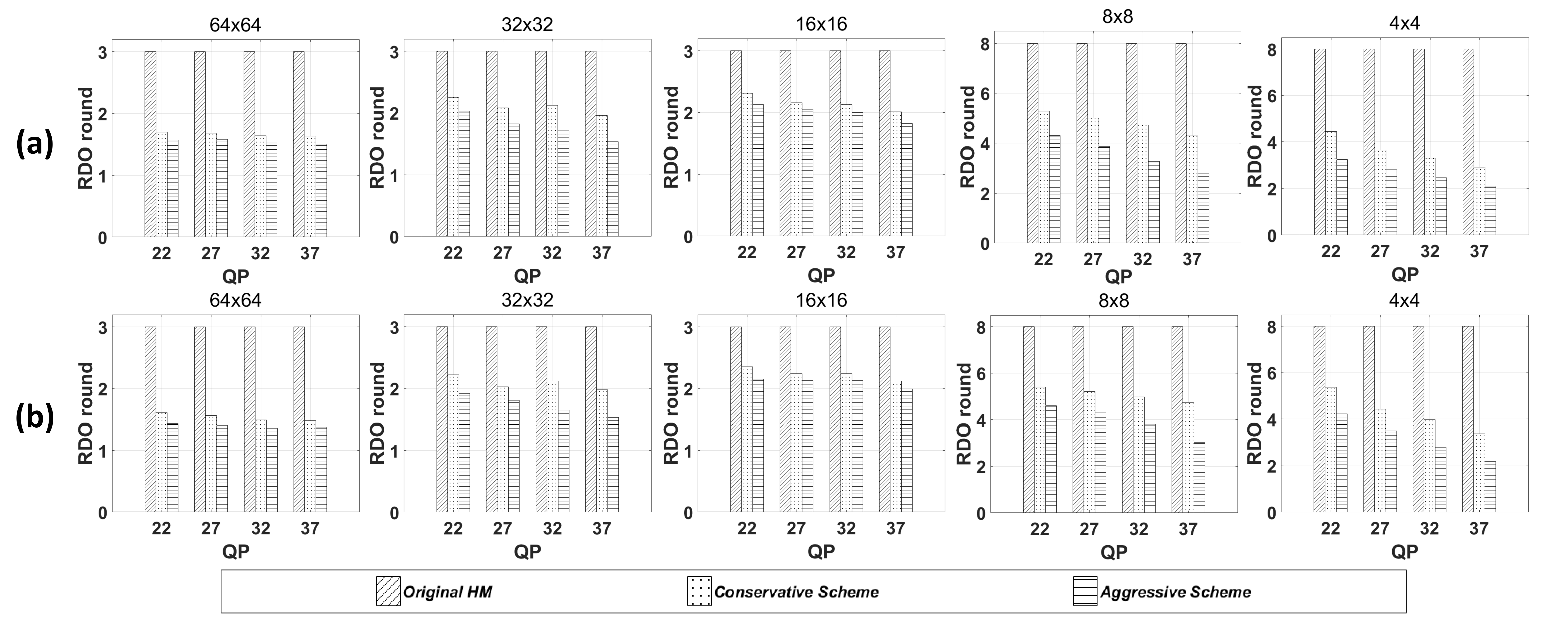}}
 	\vspace{-0cm}
 	\caption{Number of RDO modes comparisons between HM and our scheme. (a) Traffic in class A. (b) RaceHorses in class D. Slanted lines indicate HM, while dots and horizontal lines are for the conservative and aggressive schemes, respectively.}
 	\label{rdot}
 	\vspace{-0.3cm}
 \end{figure*}

\textbf{Evaluation on RDO rounds.}
First, we examine the number of modes for RDO in our approach. As we train the AK-CNNs for fast intra-mode decision in two types of settings, we apply both the conservative and aggressive schemes to two JCT-VC test sequences, Traffic in Class A and RaceHorses in Class D, and then compare it with the original HM 16.9. The average number of candidate modes for RDO (MPMs are not included) is shown in Fig. \ref{rdot}. We can find that the final number of RDO modes is curtailed for a large percent, which leads to a significant reduction in complexity. Furthermore, the number of RDO modes in every depth is different among different sequences, which is highly correlated with the content, showing the adaptiveness of our scheme.

\textbf{Evaluation in terms of complexity reduction.}
Next, we evaluate the performance of our learned fast mode decision (LFMD) at complexity reduction. We implement both the conservative and aggressive schemes on the 18 JCT-VC test sequences and then compare them with two state-of-the-art results \cite{7457241} and \cite{8412615}, as reported in Table \ref{tab-mode}. Our conservative scheme can reduce the encoding time by 22.0\%, which is better than both \cite{7457241} ($\Delta T = 16.1\%$) and \cite{8412615} ($\Delta T = 17.7\%$). As for the aggressive scheme, it curtails the number of modes for RDO for a larger percent, so the encoding time saving rate is higher, which reaches 25.0\%. This performance exceeds that of the other two methods significantly, as 8.9\% and 7.3\% encoding time is additionally saved.

Note that our approaches can reduce the complexity more than the other two methods. This is because the heuristic approaches \cite{7457241} and \cite{8412615} only consider some situations, such as homogeneous, vertical texture and horizontal texture blocks. On the contrary, our data-driven approach can classify blocks more comprehensively, reducing more modes for RDO. Therefore, our approach can improve the efficiency of HEVC intra coding.

\textbf{Evaluation on RD performance.}
Then, we use the RD performance to evaluate our approach, in terms of BD-BR. Table \ref{tab-mode} tabulates the BD-BR results, with the original HM 16.9 as the anchor. Our conservative scheme incurs a BD-BR increase of 0.18\%, better than \cite{8412615} (0.52\% increase) and \cite{7457241} (0.21\% increase), which is owing to the high precision of the MNRC mechanism and AK-CNN predictors. For our aggressive scheme, it reduces more candidates for RDO; thus, the BD-BR increase reaches 0.36\%.

More importantly, we calculate the standard deviations of the BD-BR increases of the four approaches. We can observe that both our conservative scheme ($std= 0.05\%$) and aggressive scheme ($std= 0.09\%$) have less fluctuations than \cite{8412615} ($std= 0.16\%$) and \cite{7457241} ($std= 0.13\%$) among 18 test sequences, which indicates that our CNN-based approach has better generalization capability than those two heuristic methods.

 It is also very interesting to see how our algorithm can preserve the coding performance with so many RDO candidate modes removed. Here, we record the final chosen modes in all depths with three settings: 1. original HM with the default configuration, 2. brute-force full search without the HM fast intra-mode algorithm, and 3. proposed method with the conservative scheme. Taking the full search result as reference, we report the similarities of other two experimental results in Table \ref{tab-simi}.

\begin{table}[]
	\renewcommand{\arraystretch}{1.3}
	\caption{Similarity of chosen modes with the full search scheme}
	\label{tab:HDCA_perform}
	\centering
	\begin{tabular}{lllll}\toprule
		& \multicolumn{2}{c}{HM \textit{vs} FS}          & \multicolumn{2}{c}{Ours \textit{vs} FS} \\ \hline
		& Traffic & RacHor                      & Traffic        & RacHor        \\\midrule
		\multicolumn{1}{l|}{Depth 0} & 64.5\%  & \multicolumn{1}{l|}{69.5\%} & 59.6\%         & 63.5\%        \\
		\multicolumn{1}{l|}{Depth 1} & 50.5\%  & \multicolumn{1}{l|}{46.4\%} & 48.6\%         & 44.5\%        \\
		\multicolumn{1}{l|}{Depth 2} & 47.6\%  & \multicolumn{1}{l|}{46.5\%} & 46.7\%         & 44.8\%        \\
		\multicolumn{1}{l|}{Depth 3} & 43.9\%  & \multicolumn{1}{l|}{44.4\%} & 43.6\%         & 43.6\%        \\
		\multicolumn{1}{l|}{Depth 4} & 39.9\%  & \multicolumn{1}{l|}{38.5\%} & 39.5\%         & 37.7\%    
		\\\bottomrule   
	\end{tabular}
	 	\begin{tablenotes}
	 		\item RaceHorses is abbreviated as RacHor, while FS is for full search.
	 	\end{tablenotes}
 	\vspace{-0.3cm}
 	\label{tab-simi}
\end{table}

From this table, we can see that the similarities of both original HM and our method are not very high, especially for small blocks. The main reason we believe is that different reconstructed pixels are used as reference, which is a very important factor for intra prediction. In other word, for one block, the optimal intra-mode can be different with different reconstructed pixels. Consequently, directly comparing the final chosen modes may not be an effective method. 

Here, we use \textit{cover rate} to measure the prediction effectiveness, which is defined as the ratio of PUs for which the RDO rounds are not smaller than the MNRC. This value can reveal the ability of finding the optimal mode as effective as the original HM under different reconstructed pixels. The cover rates of these two sequences are reported in Table \ref{tab-cover}. We can see that the cover rate of our scheme is rather high, which guarantees RD performance. 

\begin{table}[h]
	\renewcommand{\arraystretch}{1.3}
	\caption{Cover rate of two sequences in all depths}
	\centering
	\begin{tabular}{@{}ccc@{}}
		\toprule
		\multicolumn{1}{l}{}         & Traffic & RaceHorses \\ \midrule
		\multicolumn{1}{c|}{Depth 0}     & 77.5\% & 78.4\%     \\
		\multicolumn{1}{c|}{Depth 1}     & 87.6\% & 85.8\%     \\
		\multicolumn{1}{c|}{Depth 2}     & 90.7\% & 90.1\%     \\
		\multicolumn{1}{c|}{Depth 3}     & 96.4\% & 95.1\%    \\ 
		\multicolumn{1}{c|}{Depth 4}     & 96.1\% & 95.4\%    \\ 
		\bottomrule
	\end{tabular}
\vspace{-0.4cm}
\label{tab-cover}
\end{table}

\subsection{Performance Evaluation of the Overall Approach}
 We assemble the fast CU/PU size decision (OT mode) and the fast mode decision (conservative scheme) to obtain an overall system since these two modes achieve balanced trade-off between complexity reduction and RD performance. Then, we implement this system on the 18 JCT-VC test sequences. Table \ref{tab-mode} reports the BD-BR and time saving rate of the proposed overall system. For the RD performance, the overall system incurs 2.09\% extra BD-BR, which is negligible but still better than \cite{7547305} and \cite{8384310}. The time saving rate reaches 75.2\%, which allows for the encoder to accelerate at a 4$\times$ speed. 

\subsection{Analysis for Asymmetric Kernel}
Previous researches such as \cite{dai2017deformable} explore convolutional kernels of different shapes for better recognition. In this paper, we design the special asymmetric kernel targeting at the near-vertical and near-horizontal textures, which is essential for intra coding. To demonstrate the effectiveness, we replace the asymmetric kernels with the normal square kernels (\textit{e.g.}, 5$\times$9 to 7$\times$7). Using the same training strategy, we obtain a group of CNN models as reference. Then, we compare the performance in our validation set for fast size decision. Table \ref{tab-pred} reports the result, which is averaged over four QPs.

\begin{table}[!!h]
	\renewcommand{\arraystretch}{1.3}
	\caption{Prediction accuracy comparison}
	\centering
	\begin{tabular}{@{}ccc@{}}
		\toprule
		\multicolumn{1}{l}{}         & Asymmetric kernel & Square kernel \\ \midrule
		\multicolumn{1}{c|}{Depth 0}     & 92.51\% & 92.42\%     \\
		\multicolumn{1}{c|}{Depth 1}     & 86.03\% & 85.74\%     \\
		\multicolumn{1}{c|}{Depth 2}     & 81.87\% & 81.53\%     \\
		\multicolumn{1}{c|}{Depth 3}     & 77.43\% & 77.06\%    \\ \bottomrule
	\end{tabular}
\label{tab-pred}
\end{table}

From Table \ref{tab-pred}, we can observe that the asymmetric kernel helps get better prediction precision than the normal square kernel. With the diversified features from the multibranch of asymmetric kernels and square kernels, AK-CNN models are able to better recognize the characteristics of current block. As a consequence, the prediction performance is improved.

 \subsection{Model Complexity Analysis}	
\textbf{Computational complexity.}
Employing CNNs in our algorithm will introduce an additional computational complexity. Actually, our CNN models for both tasks consume less than 1\% of the encoding time required by the anchor of HM 16.9, and this has been taken into consideration of encoding time reduction. 

Here, we give a detailed analysis about the additional computational complexity of our CNN models via counting the number of floating-point operations, including additions and multiplications. We calculate and obtain that the number of floating-point operations of our CNN models ranges from 1.28$\times$$10^{4}$ (depth = 4) to 6.35$\times$$10^{5}$ (depth = 0) in single precision (32-bit). This result shows that our models perform much fewer floating-point operations than the common CNN models such as AlexNet \cite{krizhevsky2012imagenet} ($\sim$ 3$\times$$10^{9}$ floating-point operations) or VGG-16 \cite{simonyan2014very} ($\sim$ 4$\times$$10^{10}$ floating-point operations) by at least four orders of magnitude.		
		
For fast CU/PU size decision part, early termination mechanism is adopted, which means that if current CU/PU is not split, the inference of CNNs for the sub-CU/PU will be skipped. For the fast mode decision part, the CNN inference is only performed for the depth decided by the size decision part. As a result, a large proportion of the computational complexity of the CNNs can be reduced. The inference ratio is defined as the percentage of blocks on which we need to implement prediction. Here, we provide the inference ratio of our CNN models in all depths for two sequences, the high-definition sequence (BasketballDrive in Class B) and the low-definition sequence (RaceHorses in Class D), in Table \ref{tab-infer}.

\begin{table}[h]
	\renewcommand{\arraystretch}{1.3}
	\caption{Inference ratios of CNN models at all depths}
	\centering
	\begin{tabular}{ccccc}
		\toprule
		& \multicolumn{2}{c}{Size Decision}   & \multicolumn{2}{c}{Mode Decision} \\ \hline
		& BasDri & RacHor                     & BasDri          & RacHor          \\ \midrule
		\multicolumn{1}{c|}{Depth 0} & 100.0\%  & \multicolumn{1}{c|}{100.0\%} & 13.5\%            & 3.1\%             \\
		\multicolumn{1}{c|}{Depth 1} & 86.5\%   & \multicolumn{1}{c|}{96.9\%}  & 39.6\%            & 13.5\%            \\
		\multicolumn{1}{c|}{Depth 2} & 46.9\%   & \multicolumn{1}{c|}{83.4\%}  & 29.8\%            & 28.1\%            \\
		\multicolumn{1}{c|}{Depth 3} & 17.1\%   & \multicolumn{1}{c|}{55.3\%}  & 13.4\%            & 34.6\%            \\
		\multicolumn{1}{c|}{Depth 4} & -      & \multicolumn{1}{c|}{-}     & 3.7\%             & 20.7\%            \\ \bottomrule
	\end{tabular}
 	\begin{tablenotes}
	\item BasketballDrive and RaceHorses are abbreviated as BasDri and RacHor, respectively.
\end{tablenotes}
\vspace{-0.1cm}
\label{tab-infer}
\end{table}

More importantly, since our CNN models only make use of the pixels in current CU/PU, without the need for any intermediate features during encoding, the prediction inference can be highly parallelized. The CPU test platform of our experiments, an Intel (R) Core (TM) i7-7700K CPU, supports 5.75$\times$$10^{11}$ single-precision floating-point operations per second \cite{intel}, which exceeds the computation of our CNN models by a large margin. Thus, we are able to set a large batch size for the input blocks. Based on the highly efficient deep learning framework $TensorFlow$, such a calculation can be accelerated with the generalized matrix multiplication (GEMM) \cite{lawson1979basic}. As a result, the inference time of our CNN models is greatly reduced, making it less than 1\% of the encoding time of the anchor of HM 16.9. Compared with the large percentage of encoding time reduction, our approach introduces very little time overhead in alleviating the complexity of HEVC intra coding.

In addition, with the specified hardware support, such as GPUs or field programmable gate arrays (FPGAs), the inference time of our CNN models can be further reduced drastically. When running on our GPU platform for testing, the inference can be further accelerated hundreds of times. With increasingly many terminals equipped with such hardware, we believe that our CNN-based approach can be really competitive in the future.

\textbf{Space complexity.}
The space complexity is also very important. Thus, we need to use the shallow and thin CNNs. Actually, the neural network structure in our work is really light. For fast CU/PU size decision, the number of parameters of the AK-CNNs ranges from 4.33$\times$$10^{4}$ to 4.40$\times$$10^{4}$, which is much fewer than AlexNet (6.10$\times$$10^{7}$) and VGG-16 (1.38$\times$$10^{8}$). Table \ref{tab-param} presents a comparison of our AK-CNNs  and a similar work \cite{8019316} in terms of the number of parameters.

\begin{table}[!h]
	\vspace{-0.1cm}
	\renewcommand{\arraystretch}{1.3}
	\caption{Comparison on amount of parameters}
	\centering
	\begin{tabular}{@{}ccc@{}}
		\toprule
		\multicolumn{1}{l}{}         & Ours & \cite{8019316} \\ \midrule
		\multicolumn{1}{c|}{Depth 0}     & 43,986 & 1,384,464     \\
		\multicolumn{1}{c|}{Depth 1}     & 43,602 & 1,160,208     \\
		\multicolumn{1}{c|}{Depth 2}     & 43,346 & 1,104,144     \\
		\multicolumn{1}{c|}{Depth 3}     & 43,346 & None    \\ \bottomrule
	\end{tabular}
\label{tab-param}
\end{table}

\section{Conclusions}
In this paper, we propose a LFHI framework to reduce the complexity of HEVC intra coding while maintaining the RD performance. A novel network structure with multibranch asymmetric convolutional kernels (\textit{i.e.}, AK-CNN) is proposed to better extract texture features from blocks without too much complexity. Then, we introduce the MNRC as a new concept into the fast intra-mode decision; thus, the candidates for RDO can be reduced. The EOTD scheme is used to achieve configurable complexity-efficiency tradeoffs to meet different needs. In addition, we design an interpolation-based prediction method to deal with the problem of variant QPs. To meet the need of training data, we establish the EHIC dataset, with which offline training can be efficient. Compared with the original HM 16.9, our approach reduces the encoding time by 75.2\% with a negligible 2.09\% BD-BR increase over the JCT-VC standard test sequences. 

In future work, we will consider extending our scheme to H.266/VVC for further optimization. The partition pattern in VVC is much more complicated than HEVC, so directly applying the separate prediction scheme is not very suitable, a problem for which we will attempt to find an appropriate solution.

\ifCLASSOPTIONcaptionsoff
  \newpage
\fi
\bibliographystyle{IEEEtran}
\bibliography{IEEEexample,reference} 

\begin{thebibliography}{10}
\providecommand{\url}[1]{#1}
\csname url@samestyle\endcsname
\providecommand{\newblock}{\relax}
\providecommand{\bibinfo}[2]{#2}
\providecommand{\BIBentrySTDinterwordspacing}{\spaceskip=0pt\relax}
\providecommand{\BIBentryALTinterwordstretchfactor}{4}
\providecommand{\BIBentryALTinterwordspacing}{\spaceskip=\fontdimen2\font plus
\BIBentryALTinterwordstretchfactor\fontdimen3\font minus
  \fontdimen4\font\relax}
\providecommand{\BIBforeignlanguage}[2]{{%
\expandafter\ifx\csname l@#1\endcsname\relax
\typeout{** WARNING: IEEEtran.bst: No hyphenation pattern has been}%
\typeout{** loaded for the language `#1'. Using the pattern for}%
\typeout{** the default language instead.}%
\else
\language=\csname l@#1\endcsname
\fi
#2}}
\providecommand{\BIBdecl}{\relax}
\BIBdecl

\bibitem{6316136}
G.~J. Sullivan, J.~Ohm, W.~Han, and T.~Wiegand, ``{Overview of the High
  Efficiency Video Coding (HEVC)} standard,'' \emph{IEEE Transactions on
  Circuits and Systems for Video Technology}, vol.~22, no.~12, pp. 1649--1668,
  Dec 2012.

\bibitem{1218189}
T.~Wiegand, G.~J. Sullivan, G.~Bjontegaard, and A.~Luthra, ``{Overview of the
  H.264/AVC video coding standard},'' \emph{IEEE Transactions on Circuits and
  Systems for Video Technology}, vol.~13, no.~7, pp. 560--576, July 2003.

\bibitem{6317153}
J.~Lainema, F.~Bossen, W.~Han, J.~Min, and K.~Ugur, ``Intra coding of the
  {HEVC} standard,'' \emph{IEEE Transactions on Circuits and Systems for Video
  Technology}, vol.~22, no.~12, pp. 1792--1801, Dec 2012.

\bibitem{6983560}
K.~Lim, J.~Lee, S.~Kim, and S.~Lee, ``Fast {PU} skip and split termination
  algorithm for {HEVC} intra prediction,'' \emph{IEEE Transactions on Circuits
  and Systems for Video Technology}, vol.~25, no.~8, pp. 1335--1346, Aug 2015.

\bibitem{6470665}
S.~Cho and M.~Kim, ``Fast {CU} splitting and pruning for suboptimal {CU}
  partitioning in {HEVC} intra coding,'' \emph{IEEE Transactions on Circuits
  and Systems for Video Technology}, vol.~23, no.~9, pp. 1555--1564, Sept 2013.

\bibitem{zhao2014fast}
L.~Zhao, X.~Fan, S.~Ma, and D.~Zhao, ``Fast intra-encoding algorithm for {High
  Efficiency Video Coding},'' \emph{Signal Processing: Image Communication},
  vol.~29, no.~9, pp. 935--944, 2014.

\bibitem{7538911}
H.~Fang, H.~Chen, and T.~Chang, ``Fast intra prediction algorithm and design
  for {High Efficiency Video Coding},'' in \emph{2016 IEEE International
  Symposium on Circuits and Systems (ISCAS)}, May 2016, pp. 1770--1773.

\bibitem{7521923}
N.~Kim, S.~Jeon, H.~J. Shim, B.~Jeon, S.~Lim, and H.~Ko, ``{Adaptive
  keypoint-based CU depth decision for HEVC intra coding},'' in \emph{2016 IEEE
  International Symposium on Broadband Multimedia Systems and Broadcasting
  (BMSB)}, June 2016, pp. 1--3.

\bibitem{zhang2019statistical}
Y.~Zhang, N.~Li, S.~Kwong, G.~Jiang, and H.~Zeng, ``Statistical early
  termination and early skip models for fast mode decision in {HEVC} intra
  coding,'' \emph{ACM Transactions on Multimedia Computing, Communications, and
  Applications (TOMM)}, vol.~15, no.~3, p.~70, 2019.

\bibitem{6738325}
M.~U.~K. Khan, M.~Shafique, and J.~Henkel, ``An adaptive complexity reduction
  scheme with fast prediction unit decision for {HEVC} intra encoding,'' in
  \emph{2013 IEEE International Conference on Image Processing (ICIP)}, Sept
  2013, pp. 1578--1582.

\bibitem{6890319}
T.~Mallikarachchi, A.~Fernando, and H.~K. Arachchi, ``Efficient coding unit
  size selection based on texture analysis for {HEVC} intra prediction,'' in
  \emph{2014 IEEE International Conference on Multimedia and Expo (ICME)}, July
  2014, pp. 1--6.

\bibitem{chiang2019fast}
J.-C. Chiang, K.-K. Peng, C.-C. Wu, C.-Y. Deng, and W.-N. Lie, ``{Fast intra
  mode decision and fast CU size decision for depth video coding in 3D-HEVC},''
  \emph{Signal Processing: Image Communication}, vol.~71, pp. 13--23, 2019.

\bibitem{7457241}
T.~Zhang, M.~Sun, D.~Zhao, and W.~Gao, ``Fast intra-mode and {CU} size decision
  for {HEVC},'' \emph{IEEE Transactions on Circuits and Systems for Video
  Technology}, vol.~27, no.~8, pp. 1714--1726, Aug 2017.

\bibitem{7024895}
N.~Hu and E.~Yang, ``Fast mode selection for {HEVC} intra-frame coding with
  entropy coding refinement based on a transparent composite model,''
  \emph{IEEE Transactions on Circuits and Systems for Video Technology},
  vol.~25, no.~9, pp. 1521--1532, Sept 2015.

\bibitem{kuanar2019adaptive}
S.~Kuanar, K.~Rao, M.~Bilas, and J.~Bredow, ``Adaptive {CU} mode selection in
  {HEVC} intra prediction: A deep learning approach,'' \emph{Circuits, Systems,
  and Signal Processing}, vol.~38, pp. 5081--5102, 2019.

\bibitem{7588907}
D.~Liu, X.~Liu, and Y.~Li, ``Fast {CU} size decisions for {HEVC} intra frame
  coding based on support vector machines,'' in \emph{2016 IEEE 14th Intl Conf
  on Dependable, Autonomic and Secure Computing, 14th Intl Conf on Pervasive
  Intelligence and Computing, 2nd Intl Conf on Big Data Intelligence and
  Computing and Cyber Science and Technology Congress
  (DASC/PiCom/DataCom/CyberSciTech)}, Aug 2016, pp. 594--597.

\bibitem{duanmu2016fast}
F.~Duanmu, Z.~Ma, and Y.~Wang, ``Fast mode and partition decision using machine
  learning for intra-frame coding in {HEVC} screen content coding extension,''
  \emph{IEEE Journal on Emerging and Selected Topics in Circuits and Systems},
  vol.~6, no.~4, pp. 517--531, 2016.

\bibitem{westland2019decision}
N.~Westland, A.~S. Dias, and M.~Mrak, ``Decision trees for complexity reduction
  in video compression,'' in \emph{2019 IEEE International Conference on Image
  Processing (ICIP)}.\hskip 1em plus 0.5em minus 0.4em\relax IEEE, 2019, pp.
  2666--2670.

\bibitem{kuang2019online}
W.~Kuang, Y.-L. Chan, S.-H. Tsang, and W.-C. Siu, ``Online-learning-based
  bayesian decision rule for fast intra mode and {CU} partitioning algorithm in
  {HEVC} screen content coding,'' \emph{IEEE Transactions on Image Processing},
  2019.

\bibitem{7547305}
Z.~Liu, X.~Yu, Y.~Gao, S.~Chen, X.~Ji, and D.~Wang, ``{CU} partition mode
  decision for {HEVC} hardwired intra encoder using convolution neural
  network,'' \emph{IEEE Transactions on Image Processing}, vol.~25, no.~11, pp.
  5088--5103, Nov 2016.

\bibitem{8019316}
T.~Li, M.~Xu, and X.~Deng, ``A deep convolutional neural network approach for
  complexity reduction on intra-mode {HEVC},'' in \emph{2017 IEEE International
  Conference on Multimedia and Expo (ICME)}, July 2017, pp. 1255--1260.

\bibitem{chung2017hevc}
C.-H. Chung, W.-H. Peng, and J.-H. Hu, ``{HEVC/H. 265 coding unit split
  decision using deep reinforcement learning},'' in \emph{2017 International
  Symposium on Intelligent Signal Processing and Communication Systems
  (ISPACS)}.\hskip 1em plus 0.5em minus 0.4em\relax IEEE, 2017, pp. 570--575.

\bibitem{8384310}
M.~Xu, T.~Li, Z.~Wang, X.~Deng, R.~Yang, and Z.~Guan, ``Reducing complexity of
  {HEVC}: A deep learning approach,'' \emph{IEEE Transactions on Image
  Processing}, vol.~27, no.~10, pp. 5044--5059, Oct 2018.

\bibitem{piao2010encoder}
Y.~Piao, J.~Min, J.~Chen \emph{et~al.}, ``Encoder improvement of unified intra
  prediction,'' \emph{JCTVC-C207}, vol. 7674, pp. 568--577, 2010.

\bibitem{6201851}
W.~Jiang, H.~Ma, and Y.~Chen, ``Gradient based fast mode decision algorithm for
  intra prediction in {HEVC},'' in \emph{2012 2nd International Conference on
  Consumer Electronics, Communications and Networks (CECNet)}, April 2012, pp.
  1836--1840.

\bibitem{6466835}
M.~Zhang, C.~Zhao, and J.~Xu, ``{An adaptive fast intra mode decision in
  HEVC},'' in \emph{2012 19th IEEE International Conference on Image Processing
  (ICIP)}, Sept 2012, pp. 221--224.

\bibitem{7169266}
T.~S. Kim, M.~H. Sunwoo, and J.~Chung, ``Hierarchical fast mode decision
  algorithm for intra prediction in {HEVC},'' in \emph{2015 IEEE International
  Symposium on Circuits and Systems (ISCAS)}, May 2015, pp. 2792--2795.

\bibitem{7149261}
M.~Jamali, S.~Coulombe, and F.~Caron, ``Fast {HEVC} intra mode decision based
  on edge detection and {SATD} costs classification,'' in \emph{2015 Data
  Compression Conference}, April 2015, pp. 43--52.

\bibitem{6662471}
H.~Zhang and Z.~Ma, ``Fast intra mode decision for {High Efficiency Video
  Coding (HEVC)},'' \emph{IEEE Transactions on Circuits and Systems for Video
  Technology}, vol.~24, no.~4, pp. 660--668, April 2014.

\bibitem{7805540}
W.~Liao, D.~Yang, and Z.~Chen, ``{A fast mode decision algorithm for HEVC intra
  prediction},'' in \emph{2016 Visual Communications and Image Processing
  (VCIP)}, Nov 2016, pp. 1--4.

\bibitem{shen2013fast}
L.~Shen, Z.~Zhang, and P.~An, ``{Fast CU size decision and mode decision
  algorithm for HEVC intra coding},'' \emph{IEEE Transactions on Consumer
  Electronics}, vol.~59, no.~1, pp. 207--213, 2013.

\bibitem{7362704}
S.~Jaballah, K.~Rouis, and J.~B. Tahar, ``Clustering-based fast intra
  prediction mode algorithm for {HEVC},'' in \emph{2015 23rd European Signal
  Processing Conference (EUSIPCO)}, Aug 2015, pp. 1850--1854.

\bibitem{7051618}
D.~Zhang, Y.~Chen, and E.~Izquierdo, ``{Fast intra mode decision for HEVC based
  on texture characteristic from RMD and MPM},'' in \emph{2014 IEEE Visual
  Communications and Image Processing Conference (VCIP)}, Dec 2014, pp.
  510--513.

\bibitem{8412615}
S.~Ryu and J.~Kang, ``Machine learning-based fast angular prediction mode
  decision technique in video coding,'' \emph{IEEE Transactions on Image
  Processing}, vol.~27, no.~11, pp. 5525--5538, Nov 2018.

\bibitem{laude2016deep}
T.~Laude and J.~Ostermann, ``{Deep learning-based intra prediction mode
  decision for HEVC},'' in \emph{2016 Picture Coding Symposium (PCS)}.\hskip
  1em plus 0.5em minus 0.4em\relax IEEE, 2016, pp. 1--5.

\bibitem{HM}
JCT-VC, ``{HM software},'' \url{https://hevc.hhi.fraunhofer.de/svn/svn\_HEVC
  Software/tags/HM-16.9/}.

\bibitem{hochman1969pareto}
H.~M. Hochman and J.~D. Rodgers, ``Pareto optimal redistribution,'' \emph{The
  American Economic Review}, vol.~59, no.~4, pp. 542--557, 1969.

\bibitem{miettinen2012nonlinear}
K.~Miettinen, \emph{Nonlinear multiobjective optimization}.\hskip 1em plus
  0.5em minus 0.4em\relax Springer Science \& Business Media, 2012, vol.~12.

\bibitem{zhang2007moea}
Q.~Zhang and H.~Li, ``{MOEA/D: A multiobjective evolutionary algorithm based on
  decomposition},'' \emph{IEEE Transactions on Evolutionary Computation},
  vol.~11, no.~6, pp. 712--731, 2007.

\bibitem{sutton1998reinforcement}
R.~SUTTON, ``Reinforcement learning: An introduction,'' \emph{Adaptive
  Computation and Machine Learning}, 1998.

\bibitem{schaefer2003ucid}
G.~Schaefer and M.~Stich, ``{UCID: An uncompressed color image database},'' in
  \emph{Storage and Retrieval Methods and Applications for Multimedia 2004},
  vol. 5307.\hskip 1em plus 0.5em minus 0.4em\relax International Society for
  Optics and Photonics, 2003, pp. 472--481.

\bibitem{dang2015raise}
D.-T. Dang-Nguyen, C.~Pasquini, V.~Conotter, and G.~Boato, ``{RAISE: a raw
  images dataset for digital image forensics},'' in \emph{Proceedings of the
  6th ACM Multimedia Systems Conference}.\hskip 1em plus 0.5em minus
  0.4em\relax ACM, 2015, pp. 219--224.

\bibitem{agustsson2017ntire}
E.~Agustsson and R.~Timofte, ``Ntire 2017 challenge on single image
  super-resolution: Dataset and study,'' in \emph{The IEEE Conference on
  Computer Vision and Pattern Recognition (CVPR) Workshops}, vol.~3, 2017,
  p.~2.

\bibitem{bossen2013test}
F.~Bossen and H.~Common, ``{Test conditions and software reference
  configurations, JCT-VC Doc},'' \emph{L1100, Jan}, 2013.

\bibitem{lin2018focal}
T.-Y. Lin, P.~Goyal, R.~Girshick, K.~He, and P.~Doll{\'a}r, ``Focal loss for
  dense object detection,'' \emph{IEEE Transactions on Pattern Analysis and
  Machine Intelligence}, 2018.

\bibitem{bjontegaard2001calculation}
G.~Bjontegaard, ``{Calculation of average PSNR differences between
  RD-curves},'' \emph{VCEG-M33}, 2001.

\bibitem{abadi2016tensorflow}
M.~Abadi, P.~Barham, J.~Chen, Z.~Chen, A.~Davis, J.~Dean, M.~Devin,
  S.~Ghemawat, G.~Irving, M.~Isard \emph{et~al.}, ``Tensorflow: A system for
  large-scale machine learning,'' in \emph{12th $\{$USENIX$\}$ Symposium on
  Operating Systems Design and Implementation ($\{$OSDI$\}$ 16)}, 2016, pp.
  265--283.

\bibitem{kingma2014adam}
D.~P. Kingma and J.~Ba, ``Adam: A method for stochastic optimization,''
  \emph{arXiv preprint arXiv:1412.6980}, 2014.

\bibitem{srivastava2014dropout}
N.~Srivastava, G.~Hinton, A.~Krizhevsky, I.~Sutskever, and R.~Salakhutdinov,
  ``Dropout: a simple way to prevent neural networks from overfitting,''
  \emph{The journal of machine learning research}, vol.~15, no.~1, pp.
  1929--1958, 2014.

\bibitem{dai2017deformable}
J.~Dai, H.~Qi, Y.~Xiong, Y.~Li, G.~Zhang, H.~Hu, and Y.~Wei, ``Deformable
  convolutional networks,'' in \emph{Proceedings of the IEEE international
  conference on computer vision}, 2017, pp. 764--773.

\bibitem{krizhevsky2012imagenet}
A.~Krizhevsky, I.~Sutskever, and G.~E. Hinton, ``Imagenet classification with
  deep convolutional neural networks,'' in \emph{Advances in neural information
  processing systems}, 2012, pp. 1097--1105.

\bibitem{simonyan2014very}
K.~Simonyan and A.~Zisserman, ``Very deep convolutional networks for
  large-scale image recognition,'' \emph{arXiv preprint arXiv:1409.1556}, 2014.

\bibitem{intel}
L.~Corporation, ``{Intel Core X Series i9-7900X and i7-7740X},''
  \url{https://lanoc.org/review/cpus/7566-intel-core-x-series-i9-7900x-and-i7-7740x?showall=&start=2}.

\bibitem{lawson1979basic}
C.~L. Lawson, R.~J. Hanson, D.~R. Kincaid, and F.~T. Krogh, ``Basic linear
  algebra subprograms for fortran usage,'' \emph{ACM Transactions on
  Mathematical Software (TOMS)}, vol.~5, no.~3, pp. 308--323, 1979.

\end{thebibliography}

\end{document}